\documentclass[journal,twoside]{IEEEtran}
\usepackage{amsmath}
\usepackage{amsfonts}
\usepackage{amssymb}
\usepackage{mathrsfs}
\usepackage{amsthm}
\usepackage{graphicx}
\usepackage{epstopdf}
\usepackage{caption}
\usepackage{algorithm}
\usepackage{algorithmic}
\usepackage{multirow}
\usepackage{cases}
\usepackage{color}
\usepackage{subfigure}
\usepackage{cite}
\usepackage{enumerate}

\newtheorem{lemma}{Lemma}
\newtheorem{proposition}{Proposition}

\allowdisplaybreaks[0]
\begin{document}

\title{\begin{center}Pilot Spoofing Attack by Multiple Eavesdroppers\end{center}
\author{\hspace{0.2in} Ke-Wen Huang, \hspace{0.1in} Hui-Ming Wang, \hspace{0.1in}\emph{Senior Member, IEEE},
\hspace{0.1in} Yongpeng Wu, \hspace{0.1in} \emph{Senior Member, IEEE}, \hspace{0.1in} and ~Robert Schober, \emph{Fellow, IEEE}
}
\thanks{K.-W. Huang and H.-M. Wang are with the School of Electronics and Information Engineering, and also with the MOE Key Laboratory for Intelligent Networks and Network Security, Xi'an Jiaotong University, Xi'an, 710049, China. Email: {\tt xjtu-huangkw@outlook.com, xjbswhm@gmail.com}.
}
\thanks{Y. Wu is with the Department of Electrical Engineering,  Shanghai Jiao Tong University, Minhang 200240, China. E-mail: {\tt yongpeng.wu@sjtu.edu.cn}.}
\thanks{R. Schober is with the Institute for Digital Communications, Friedrich-Alexander-University Erlangen-N\"{u}rnberg (FAU), Erlangen, Germany. E-mail: {\tt robert.schober@fau.de}.}
}
\maketitle

\begin{abstract}
In this paper, we investigate the design of a pilot spoofing attack (PSA) carried out by multiple single-antenna eavesdroppers (Eves) in a downlink time-division duplex (TDD) system, where a multiple antenna base station (BS) transmits confidential information to a single-antenna legitimate user (LU).
During the uplink channel training phase, multiple Eves collaboratively impair the channel acquisition of the legitimate link, aiming at maximizing the wiretapping signal-to-noise ratio (SNR) in the subsequent downlink data transmission phase.
Two different scenarios are investigated: (1) the BS is unaware of the PSA, and (2) the BS attempts to detect the presence of the PSA.
For both scenarios, we formulate wiretapping SNR maximization problems.
For the second scenario, we also investigate the probability of successful detection and constrain it to remain below a pre-designed threshold.
The two resulting optimization problems can be unified into a more general non-convex optimization problem, and we propose an efficient algorithm based on the minorization-maximization (MM) method and the alternating direction method of multipliers (ADMM) to solve it.
The proposed MM-ADMM algorithm is shown to converge to a stationary point of the general problem.
In addition, we propose a semi-definite relaxation (SDR) method
as a benchmark to evaluate the efficiency of the MM-ADMM algorithm.
Numerical results show that the MM-ADMM algorithm achieves near-optimal performance and is computationally more efficient than the SDR-based method.
\end{abstract}


\section{Introduction}
Physical layer security (PLS) techniques have attracted significant attention as a viable option for securing wireless communications \cite{WynerWiretap,M.Bloch2011,WangBook}.
Recently, due to the spatial degrees of freedom offered by multiple antennas, multiple-input multiple-output (MIMO) techniques  have been exploited to further enhance PLS
\cite{A.Khisti20101,A.Khisti20102,Y.-W.P.Hong2013}.
In particular, secure beamforming and artificial-noise-aided transmission are two well-known approaches to facilitate PLS
that have been considered in the context of point-to-point multiple antenna systems \cite{Y.Wu2012,Q.Li2013,S.Loyka2015,H.M.Wang2015Jan,H.-M.Wang2015}, multi-user multiple antenna systems \cite{Q.Li2011,M.F.Hanif2014,K.Cumanan2016}, and multiple relay systems \cite{H.-M.Wang2012,C.Jeong2012,C.Wang2015}.
To enhance the secrecy capacity/rate in PLS, knowledge of the channel state information (CSI) of the legitimate receiver at the transmitter is crucial.
In practice, the CSI has to be obtained by transmitting a training sequence during a training phase.
However, in most of the existing literature on PLS, the training phase has been ignored and the CSI at the transmitter is modelled as perfect \cite{Q.Li2013,S.Loyka2015} or imperfect \cite{J.Huang2012,Q.Li2011_2}.
A few works on PLS consider both the training and data transmission procedure, but are only focused on passive Eves, i.e., the Eves keep silent during both channel training and data transmission \cite{J.Zhu2014Sep,J.Zhu2016May,H.-M.Wang2016Nov,H.-M.Wang2015}.

Recently, it has been shown in \cite{X.Zhou2012} that an intelligent active eavesdropper can greatly enhance its wiretapping capability by implementing a pilot spoofing attack (PSA).
More specifically, in a time division duplex (TDD) system with a multiple-antenna base station (BS) and a single-antenna user, the downlink time slot is usually divided into two phases.
The first phase is used for uplink training where the legitimate user (LU) transmits a pilot sequence to the BS for channel estimation.
In the second phase, i.e., the downlink data transmission phase, the estimated uplink channel is regarded as the downlink channel by exploiting reciprocity, and beamforming based on this CSI is used to transmit the confidential message to the LU.
However, if an eavesdropper (Eve) attacks the uplink training phase by transmitting the same pre-designed training sequence as the LU, the estimated channel obtained at the BS is a weighted combination of the legitimate channel and the wiretap channel.
Based on this incorrect CSI, the beam formed by the BS will be oriented towards both the LU and the Eve, which results in severe signal leakage to Eve.

A number of works focused on combating the PSA \cite{D.Kapetanovic2013,S.Im2013,Q.Xiong2015,Q.Xiong2016,Wu2016TIT,Y.O.Basciftci2015,Y.O.BasciftciOnline}.
In \cite{D.Kapetanovic2013}, to detect the PSA, the authors proposed a random training scheme, wherein the training signal is randomly chosen from a set of phase-shift keying symbols.
The PSA detection probability of this scheme can approach $1$ arbitrarily close if the number of antennas at the BS is sufficiently large.
However,  the training sequence has to be transmitted
twice, which decreases spectrum  efficiency.
In \cite{S.Im2013,Q.Xiong2015}, the authors formulated the PSA detection as a binary hypothesis testing problem, and the likelihood ratio based on the energy of the received signal was used as the detection statistic.
In \cite{Q.Xiong2016}, the authors proposed a two-way training scheme to detect the PSA. If the PSA is detected successfully, the BS will simultaneously estimate the channels of the LU and Eve, and the estimates are used to safeguard  transmission via secure beamforming.
In \cite{Wu2016TIT}, the authors studied the PSA for a  multiple cell multiple user massive MIMO system.
To facilitate secure transmission, matched filter precoding combined with artificial noise generation is adopted at the BS, and the optimal power allocation policy for the signal and the artificial noise is derived.
In \cite{Y.O.Basciftci2015,Y.O.BasciftciOnline}, the authors investigated
the secure degrees of freedom (DoF) in multiple user massive MIMO systems in the presence of a full-duplex Eve who is capable of eavesdropping and jamming simultaneously. The authors proposed a scheme that hides the pilot signal assignments from Eve by utilizing
an extended pilot signal set. As a consequence, the obtained secure DoF are equal to the secure DoF when Eve does not attack.

Although the above works make important steps towards overcoming the PSA, they assume that there is only one Eve whose transmit power is fixed regardless of the CSI.
However, in practice, there may be multiple cooperating Eves
employing more intelligent methods to perform the PSA.
This is the main motivation for this paper.
In this paper, we take the point of view of the Eves, and investigate how multiple Eves can cooperatively design the PSA to achieve better wiretapping performance. We assume that the Eves know their own CSI, which allows them to adjust and optimize their attacking signals accordingly.
The consideration of multiple Eves allows us to provide a more comprehensive evaluation of the potential secrecy threats in wireless communication systems.
The main contributions of this paper are summarized as follows:

\begin{enumerate}
\item[1)] We establish a new PSA model for TDD systems wherein multiple collusive Eves collaborate to improve their wiretapping capability. Based on this model, we assume that the Eves perform the PSA collaboratively during the uplink channel training phase in order to improve the receiving signal-to-noise ratio (SNR) of a target Eve during the subsequent downlink data transmission phase. We consider two different scenarios, i.e., (a) the BS is unaware of the PSA and thus directly transmits the data after estimating the legitimate channel, and (b) the BS carries out a detection operation after each uplink channel training phase to determine whether a PSA has occurred. For both scenarios, wiretapping SNR maximization problems are formulated.

\item[2)] For the second scenario, we first investigate the successful detection probability, i.e., the probability that the PSA is successfully detected by the BS. In particular, the successful detection probability is derived under two assumptions regarding the BS's prior knowledge of the Eves' channel. In order to conceal the PSA from the BS, we assume the Eves try to keep the successful detection probability below an acceptable threshold. This leads to a corresponding constraint in the wiretapping SNR maximization problem. Thereby, from the Eves' point of view, the successful detection probability specifies the risk of being discovered. Therefore, the formulated optimization problem allows the Eves to adjust the trade-off between improving their wiretapping SNR and reducing the risk of being detected.

\item[3)] The two formulated maximization problems are unified into one general non-convex optimization problem, and we develop an efficient algorithm to solve it. To this end, we first transform the non-convex problem into a series of convex problems by invoking the minorization-maximization (MM) principle \cite{MMmethod1,MMmethod2}. Subsequently, the alternating direction method of multipliers (ADMM) \cite{S.Boyd2011,D.P.Bertisekas1989,K.Huang2016,E.ChenTCOM2017} is used to decompose the obtained convex problems into several sub-problems that either have a closed-form solution or can be efficiently solved by Newton's method. The resulting MM-ADMM algorithm is shown to converge to a stationary point of the original optimization problem. We also provide an alternative method based on semidefinite relaxation (SDR) to solve a special case of the considered general non-convex problem, along with a sufficient condition for when this method achieves the global optimum. Numerical results show that the proposed  MM-ADMM algorithm achieves near-optimal performance but requires a much lower computational complexity than the widely used SDR-based method.
\end{enumerate}

The rest of this paper is organized as follows:
In Section II, we present the system model with multiple Eves carrying out the PSA.
In Section III, for the case where the BS is unaware of the PSA, we formulate an optimization problem for maximizing the SNR of a target Eve.
In Section IV, for the case where the BS attempts to detect the presence of the PSA, we derive the successful detection probability and formulate an optimization problem for maximizing the SNR of the target Eve while keeping the probability of successful detection  under a pre-defined threshold.
In Section V, we develop the proposed MM-ADMM algorithm to solve the optimization problems established in Section III and IV. In Section VI, we present numerical results. Finally, in Section VII, we conclude the paper.

\emph{Notation:} $(\cdot)^{T}$, $(\cdot)^{*}$, $(\cdot)^{H}$, $(\cdot)^{-1}$, and $\mathrm{Tr}\left(\cdot\right)$ represent  transpose, conjugate, conjugate transpose, inverse, and trace, respectively.
$\pmb{I}_M$ denotes a $M\times M$ identity matrix.
$\mathbb{P}\left\{\cdot\right\}$ and $\mathbb{E}(\cdot)$ denote the probability and mathematical expectation, respectively.
$||\cdot||$ denotes the $l_2$ norm.
$\mathbb{C}^{N\times M}$ and $\mathbb{R}^{N\times M}$ denote
the spaces of all $N\times M$ matrices with complex-valued and real-valued elements, respectively.
$\left\langle \pmb{x},\pmb{y}\right\rangle = \pmb{x}^H\pmb{y}$ denotes the inner product.
$\mathbb{CN}(\pmb{\mu},\pmb{\Sigma} )$ and $\mathbb{N}(\pmb{\mu},\pmb{\Sigma} )$ denote the
distributions of complex and real Gaussian random vectors, respectively, with mean $\pmb{\mu}$ and covariance matrix
$\pmb{\Sigma}$. $\Gamma_S\triangleq\int_{0}^{\infty}t^{S-1}\mathrm{e}^{-t}\mathrm{d}t$, $\Gamma_S\left(x\right)\triangleq\int_{x}^{\infty}t^{S-1}\mathrm{e}^{-t}\mathrm{d}t$, and $\gamma_S\left(x\right)\triangleq\int_{0}^{x}t^{S-1}\mathrm{e}^{-t}\mathrm{d}t$ denote the Gamma, the upper incomplete Gamma, and the lower incomplete Gamma functions, respectively \cite{Book:ISG2007}.
$\Phi\left(x\right)\triangleq \frac{2}{\sqrt{\pi}}\int_0^{x}\mathrm{e}^{-t^2}\mathrm{d}t$ is the error function \cite[Eqn. 8.250.1]{Book:ISG2007}.

\begin{figure*}[!t]
	\centering
	\includegraphics[width=4.5 in]{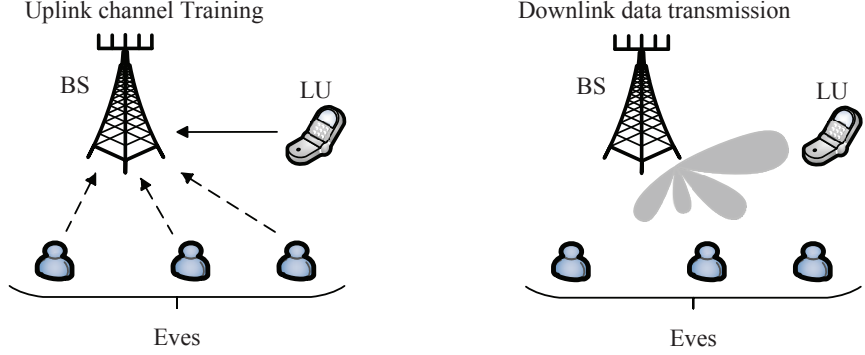}
	\caption{Multiple Eves attack the training phase by transmitting pre-designed sequences to the BS.}\label{A}
\hrulefill
\vspace{-5 mm}
\end{figure*}

\section{System Model}
We consider a TDD system where multiple single antenna Eves, who know their own CSIs, aim to intercept the signal transmitted by a multiple antennas BS to a single antenna LU by performing the PSA, as illustrated in Fig. \ref{A}.
The considered single-LU system can model a time division multiple access (TDMA) system with multiple users. More specifically, assume that the BS divides the time into several time slots having equal and fixed length. In each time slot, only one user is scheduled and the BS serves the multiple users one-by-one. Then, the user scheduled in a given time slot is the LU whereas the users scheduled in other time slots may act as Eves and attempt to intercept the data intended for the user being currently served. Therefore, for simplicity and clarity, we adopt a similar system model as in \cite{X.Zhou2012,D.Kapetanovic2013,S.Im2013,Q.Xiong2015,Q.Xiong2016}, i.e., we consider only one time slot with one LU \footnote{
In more sophisticated multi-user TDMA systems,
the lengths of different time slots may not be equal and fixed. More specifically, the times for serving different users may be dynamically optimized according to the SNRs of the users to achieve certain objectives. For example, the objective may be to achieve fairness among the users or to satisfy different quality of service requirements of the users.
In this case, the PSA affects not only the SNRs at the LU and Eves but also the optimal time allocation.
Note that the total amount of information that can be intercepted by the Eves depends  on both the wiretapping SNR and the length of the time slot allocated to the LU. Evaluating the eavesdropping capability for this case is much more complicated than for the case considered in this paper. In fact, compared to existing works such as \cite{X.Zhou2012,D.Kapetanovic2013,S.Im2013,Q.Xiong2015,Q.Xiong2016} where only one Eve is considered, in this paper, we focus on the impact of multiple Eves. Further extensions to cases where the BS adopts elaborated user scheduling schemes are left for future work.}.

The legitimate transmission procedure comprises the uplink training phase and the downlink data transmission phase.
In the uplink training phase, the LU sends a pilot sequence to the BS for channel estimation. In the data transmission phase, beamforming based on the estimated channel at the BS is utilized to transmit data to the LU.
Multiple Eves attack the legitimate link in the uplink training phase by transmitting pre-designed sequences to the BS. They aim to impair the CSI acquisition of the BS so that the downlink beamforming will be directed not only towards the LU but also towards the Eves.
In this paper, we assume that all the Eves transmit the same pilot sequence as the LU when carrying out the PSA. Note that, in general, the Eves can transmit any arbitrary pre-designed sequences instead of the pilot sequence, of course.
However, in Remark 1, we will show later that transmitting other sequences does not offer any advantage as far as intercepting the data is concerned.

We use $\pmb{h}_B\in \mathbb{C}^{N\times 1}$ and $\pmb{h}_{E,k}\in \mathbb{C}^{N\times 1}$ to denote the channel between the BS and the LU and the channel between the BS and the $k^{\mathrm{th}}$ Eve, respectively, where $N$ is the number of antennas at the BS and all channel coefficients are independent and identical distributed (i.i.d.) Gaussian random variables, i.e.,  $\pmb{h}_B\sim\mathbb{CN}\left(\pmb{0},\pmb{I}_N\right)$ and $\pmb{h}_{E,k}\sim\mathbb{CN}\left(\pmb{0},\pmb{I}_N\right)$.
In the uplink training phase, the LU and the Eves send the same pilot sequence $\pmb{x}$ to the BS. The received signal at the BS, denoted by $\pmb{Y}_T\in\mathbb{C}^{N\times \tau}$, is given by
\begin{align}\label{Training}
	\pmb{Y}_T = \sqrt{P_T}\pmb{h}_B\pmb{x}^{T}+\sum\nolimits_{k=1}^{K}\nu_k\pmb{h}_{E,k}\pmb{x}^T+\pmb{U},
\end{align}
where the training sequence $\pmb{x}\in \mathbb{C}^{\tau\times 1}$ of length $\tau$ satisfies $\pmb{x}^{T}\pmb{x}^*=\tau$, $P_T$ is the training power of the LU, $\nu_k$, for $k=1,2,\cdot\cdot\cdot,K$, are the complex weight coefficients of the $k^{\mathrm{th}}$ Eve, which will be optimized based on the Eves' CSIs to improve the wiretapping capability; $\pmb{U}\in\mathbb{C}^{N\times\tau}$ is the additive Gaussian white noise (AWGN) matrix at the BS
with each element being distributed as $\mathbb{CN}\left(0,\sigma_T^2\right)$.

To estimate the channel of the LU, $\pmb{h}_B$, the BS applies the following transformation
\begin{align}
\label{ReceivedPilot}
	\pmb{y}_T \triangleq \frac{1}{\tau\sqrt{P_T}}\pmb{Y}_T\pmb{x}^*=\pmb{h}_B+\sum_{k=1}^{K}\frac{\nu_k}{\sqrt{P_T}}\pmb{h}_{E,k}+\pmb{z},
\end{align}
where $\pmb{z}\triangleq \frac{1}{\tau\sqrt{P_T}}\pmb{U}\pmb{x}^*$ is the equivalent noise vector with distribution  $\mathbb{CN}\left(\pmb{0},\frac{\sigma_T^2}{\tau P_T}\pmb{I}_N\right)$.
Then, based on $\pmb{y}_T$, the BS estimates $\pmb{h}_B$ by using the linear minimum mean square error (LMMSE) method, which yields
\begin{align}
	\hat{\pmb{h}}_B = \pmb{L}^H\pmb{y}_T,
\label{estimation}
\end{align}
where $\pmb{L}$ is the LMMSE estimation matrix.
The construction of $\pmb{L}$ will be discussed later.
After obtaining the estimated channel, i.e., $\hat{\pmb{h}}_B$,  the BS transmits the confidential message $s$ in the downlink data transmission phase using maximal-ratio transmission (MRT) with beamforming vector $\pmb{w} \triangleq \frac{\hat{\pmb{h}}_B}{\left\|\hat{\pmb{h}}_B\right\|}$ \cite{Q.Xiong2015,Q.Xiong2016}.
The signals received by the LU and the $k^{\mathrm{th}}$ Eve are denoted by $y_{LU}$ and $y_{E,k}$, respectively, and can be written as
\begin{align}
{{y}}_{LU}& = \pmb{{h}}_B^H\frac{\hat{\pmb{h}}_B}{\|\hat{\pmb{h}}_B\|} s + n_{LU}, \  {y}_{E,k}= \pmb{{h}}_{E,k}^H\frac{\hat{\pmb{h}}_B}{\|\hat{\pmb{h}}_B\|} s + n_{E,k},\label{signalreceivedEveandLU}
\end{align}
where $n_{LU}\sim \mathbb{CN}\left(0,\sigma_{LU}^2\right)$ and $n_{E,k}\sim \mathbb{CN}\left(0,\sigma_{E,k}^2\right)$ are the additive noise at the LU and the $k^{\mathrm{th}}$ Eve, respectively.

Due to the PSA, the obtained $\hat{\pmb{h}}_B$ is a function of both ${\pmb{h}}_B$ and ${\pmb{h}}_{E,k}, k=1,2,\cdots, K$, which has two negative effects: on the one hand, the MRT beamforming vector will not  match the legitimate channel ${\pmb{h}}_B$, which leads to a power attenuation for the LU; on the other hand, each Eve will receive a copy of the leaked confidential signal.
From the point of view of the Eves, the PSA aims to maximize the received SNR of the confidential signal at the Eves to achieve a better wiretapping performance.

{In this paper, we consider the case where the confidential information has to be decoded in an online manner. Furthermore, we assume that the Eves are connected to each other via low-cost low-capacity wireless links.
Hence, we assume that the Eves can share their CSIs for performing the PSA but are not capable of sharing the received information-carrying signals to perform joint information decoding, i.e., the wireless links between the Eves provide enough capacity for sharing their CSIs but not enough capacity for sharing the real-time information-carrying signals.
{For example, the Eves may be distributed nodes which form a temporary low-cost wireless ad hoc network to communicate with each other.
In a wireless ad hoc network, the wireless links between different pairs of the nodes usually have different capacities, and for some node pairs, the links between them may have very low capacity due to shadowing and the long-distance transmission.
As a result, the CSIs of the Eves may be successfully shared as the information content of the Eves' CSI is generally small, but it may not be possible to share the received information-carrying signals at the Eves, which are usually of high  rate, to perform joint decoding.}
In this context, we also note that the more information the Eves exchange, the easier it will be for the legitimate system to detect their presence.
Therefore, in this paper, we assume that the Eves cooperate with each other to perform the PSA during the channel training phase to maximize the SNR at one of the Eve, referred to as the \emph{target Eve}, in the subsequent data transmission phase.}
The target Eve performs decoding to obtain the confidential information\footnote{
If the multiple Eves are connected via  high-cost high-capacity backhaul links or are co-located (i.e., a single Eve with multiple antennas), then they can share their received signals and perform joint information decoding based on all received signals.
In this case, the Eves can jointly design the PSA signals during the channel training phase and the combining weights during the data transmission phase, which will lead to an improved interception performance. However, this also substantially affects and complicates the wiretapping SNR maximization problem. Hence, considering this case is beyond the scope of this paper, but constitutes an interesting topic for future research.
}.
Without loss of generality, we take the $K^{\mathrm{th}}$ Eve as the target Eve.
More specifically, we formulate an optimization problem for the weight coefficients $\nu_k, k=1,2,\cdots,K$, in \eqref{ReceivedPilot} to maximize the SNR of the $K^{\mathrm{th}}$ Eve.
The resulting wiretapping performance can be regarded as a lower bound on the performance when the Eves can perform joint information decoding.

Depending on the capabilities of the BS,  the following two cases may be distinguished:
\begin{itemize}
\item  The BS is completely unaware of the existence of the PSA, i.e., the BS assumes that only the LU is transmitting the pilot sequence during the training phase.

\item The BS is cautious and after each uplink channel training phase, a detection operation is executed to determine whether a PSA has occurred. In this case, the Eves have to design the PSA more carefully.
\end{itemize}
In the following sections, we discuss these two cases and present the corresponding optimization problems one by one.

\emph{Remark 1:} If the Eves transmit any other sequence $\pmb{q}_k, k=1,2,\cdots, K,$ instead of $\pmb{x}$ in \eqref{Training},  then in the uplink training phase, $\pmb{y}_T$ in \eqref{ReceivedPilot} becomes $\tilde{\pmb{y}}_T = \pmb{h}_B + \sum_{k=1}^{K}\frac{\pmb{q}_k^T\pmb{x}^*}{\tau\sqrt{P_T}}\pmb{h}_{E,k} +\pmb{z}$. It is obvious that, as long as $\pmb{q}_k^T\pmb{x}^*\neq0,~k=1,2,\cdot\cdot\cdot,K$, holds, $\tilde{\pmb{y}}_T$ is equivalent to $\pmb{y}_T$ if we let $\nu_k = \frac{\pmb{q}_k^T\pmb{x}^*}{\tau}$. Therefore, we assume that all Eves transmit $\pmb{x}$ during the channel training phase for simplicity.

\section{PSA design when the BS is unaware of the PSA}
In this section, we consider the case when the BS is completely unaware of the PSA. In this scenario, the BS regards the received pilot signals in the uplink training phase as  the legitimate channel coefficients polluted by noise.
More specifically, the BS will  mistake $\pmb{y}_T = \pmb{h}_B+\sum_{k=1}^{K}\frac{\nu_k}{\sqrt{P_T}}\pmb{h}_{E,k}+\pmb{z}$ for $\pmb{y}_T = \pmb{h}_B + \pmb{z}$ by ignoring the second term on the right hand side of \eqref{ReceivedPilot}, and estimate $\pmb{h}_B$ according to the LMMSE principle by setting $\pmb{L}=\frac{\tau P_T}{\tau P_T + \sigma_T^2}\pmb{I}_N$ in \eqref{estimation} \cite{BOOK:S.M.Kay}. Then, the estimated channel is given by $
 \hat{\pmb{h}}_B = \alpha\left(\pmb{h}_B+\sum_{k=1}^{K}\frac{\nu_k}{\sqrt{P_T}}\pmb{h}_{E,k}+\pmb{z}\right) =
 \alpha\left(\pmb{h}_B+\pmb{h}_E+\pmb{z}\right)$,
where $\alpha \triangleq \frac{\tau P_T}{\tau P_T + \sigma_T^2}$ and $\pmb{h}_E\triangleq\sum_{k=1}^{K}\frac{\nu_k}{\sqrt{P_T}}\pmb{h}_{E,k}$.
Inserting $\hat{\pmb{h}}_B$ into \eqref{signalreceivedEveandLU}, the signal received by the LU during the signal transmission phase can be written as
\begin{align}
	{y}_{LU}& = \underbrace{\|\hat{\pmb{h}}_B\|s}_{\text{desired signal of the LU}} + \quad \underbrace{\tilde{\pmb{{h}}}_B^H\frac{\hat{\pmb{h}}_B}{\|\hat{\pmb{h}}_B\|} s + n_{LU}}_{\text{equivalent noise of the LU}}, \label{signalreceivedLU2}
\end{align}
where $n_{LU}\sim \mathbb{CN}\left(0,\sigma_L^2\right)$ is the Gaussian noise, $\tilde{\pmb{{h}}}_B$ is the channel estimation error satisfying $\pmb{{h}}_B = \tilde{\pmb{{h}}}_B + \hat{\pmb{h}}_B$ which is not known by the BS.
In practice, when there is no attack, $\tilde{\pmb{{h}}}_B$ is generally very small and thus, only the first term in \eqref{signalreceivedLU2} is regarded as the desired signal of the LU and the other terms are considered as the equivalent noise.
To decode the signal, the LU has to know the value of $\|\hat{\pmb{h}}_B\|$, which could be achieved by direct feedback from the BS.

According to the general principle of PLS, we do not assume that the LU has CSI knowledge that the Eves do not have. Therefore, we
assume that the Eves will also know $\|\hat{\pmb{h}}_B\|$ during the signal transmission phase.
{ In fact, the Eves can acquire this knowledge when the BS feeds back $\|\hat{\pmb{h}}_B\|$ to the LU.}
Then, the received signal of the $K^{\mathrm{th}}$ Eve in \eqref{signalreceivedEveandLU} can be rewritten as
\begin{align}
&\breve{y}_{E,K} =\|\hat{\pmb{h}}_B\|{y}_{E,K}\nonumber \\
& =
	\underbrace{\alpha\pmb{h}_{E,K}^H\pmb{h}_Es}_{\text{desired signal}} +
    \underbrace{\alpha\pmb{h}_{E,K}^H\pmb{h}_Bs +\alpha\pmb{h}_{E,K}^H\pmb{z}s+\|\hat{\pmb{h}}_B\|n_{E,K}}_{\text{equivalent noise $\breve{n}_{E,K}$}}.
    \label{EquivalentyEK}
\end{align}
In the following, we analyze the SNR of the $K^{\mathrm{th}}$ Eve.

\subsection{The SNR of $K^{\mathrm{th}}$ Eve}
The first three terms on the right hand side of \eqref{EquivalentyEK} contain the signal $s$, and
the desired signal of the Eves depends on their prior knowledge of $\pmb h_B$.
In general, it is difficult for the Eves to know the exact $\pmb{h}_B$, and thus only the first term in \eqref{EquivalentyEK} is regarded as the desired signal for the $K^{\mathrm{th}}$ Eve and the remaining terms are treated as equivalent noise, denoted by $\breve{n}_{E,K}$.
The equivalent noise power can be calculated as
\begin{align}
\mathbb{E}\left(\left|\breve{n}_{E,K}\right|^2\right)&=\alpha^2\bigg(P_S\sigma_{BT}^2\left\|\pmb{h}_{E,K}\right\|^2 \nonumber \\
&\quad + N\sigma_{BT}^2\sigma_{E,K}^2 + \left\|\pmb{h}_E\right\|^2\sigma_{E,K}^2\bigg),
\label{PowerEquivalentNoise}
\end{align}
where $P_S\triangleq\mathbb{E}\left(\left|s\right|^2\right)$ is the signal power transmitted by the BS,  and $\sigma_{BT}^2\triangleq\left(1+\frac{\sigma_T^2}{\tau P_T}\right)$.
Note that the expectation operation in \eqref{PowerEquivalentNoise} is with respect to (w.r.t.) $\pmb{h}_B$
, $\pmb{z}$, $s$, and $n_{E,K}$ because all of these variables are unknown to the Eves, and thus are treated as random
\footnote{
Although we have assumed in \eqref{EquivalentyEK} that the Eves know $\|\hat{\pmb{h}}_B\|$ for decoding, they do not know $\|\hat{\pmb{h}}_B\|$ for designing the PSA signals. This is because $\hat{\pmb{h}}_B$ is the estimated CSI at the BS and can only be obtained after the training phase.
The design of the PSA signals has to occur before the training phase, and at this time, the Eves do not know $\|\hat{\pmb{h}}_B\|$ in \eqref{EquivalentyEK}.
}.
Therefore, an \emph{achievable average SNR} of the $K^{\mathrm{th}}$ Eve can be written as
\begin{align}
	&\mathrm{SNR}_{E,K} \nonumber \\
	=& \frac{P_S\left|\pmb{h}_{E,K}^H\pmb{h}_E\right|^2}{
		P_S\sigma_{BT}^2\left\|\pmb{h}_{E,K}\right\|^2 +
		N\sigma_{BT}^2\sigma_{E,K}^2+\left\|\pmb{h}_E\right\|^2\sigma_{E,K}^2}.
\label{SNREk}
\end{align}

{
In practice, the Eves can not obtain the exact CSI of the LU. Nevertheless, it is of interest to consider the case where the Eves know $\pmb{h}_B$ perfectly, because the corresponding SNR constitutes an upper bound on the SNR when $\pmb{h}_B$ is not or imperfectly known to the Eves.
In fact, in practice, the LU's CSI is highly dependent on the LU's location. If the LU's location is known to the Eves, they may be able to infer partial knowledge of the LU's CSI, which means the actual eavesdropping performance is expected to be between the performance when the Eves know $\pmb{h}_B$ perfectly and the performance when the Eves do not have any knowledge of $\pmb{h}_B$.
Hence, considering the case where the Eves know $\pmb{h}_B$ perfectly yields an upper bound on the maximum performance that  the Eves can achieve by performing the PSA. Note that similar assumptions have been made in several existing works, e.g., \cite{Q.Liu2015,Z.H.Awan2012}. With this pessimistic assumption for the LU, both the first and second terms on the right hand side of \eqref{EquivalentyEK} constitute the desired signal for the $K^{\mathrm{th}}$  Eve.
In this case, an achievable average SNR is given by
\begin{align}
	&\overline{\mathrm{SNR}}_{E,K}\nonumber\\
	= &\frac{P_S\left|\pmb{h}_{E,K}^H\pmb{h}_{E} + \pmb{h}_{E,K}^H\pmb{h}_B\right|^2}{
		\frac{P_S\sigma_T^2}{\tau P_T}\left\|\pmb{h}_{E,K}\right\|^2 +
		N\frac{\sigma_T^2}{\tau P_T}\sigma_{E,K}^2+\left\|\pmb{h}_E+\pmb{h}_B\right\|^2\sigma_{E,K}^2}.\label{SNREkUpperBound}
\end{align}
}

\subsection{Design of the Attacking Scheme}
According to the discussions in Section II, the goal of the PSA is to maximize the SNR of the $K^{\mathrm{th}}$ Eve.
Based on \eqref{SNREk} and \eqref{SNREkUpperBound}, we establish the following optimization problem
\begin{align}
\label{TheSecondProblem}
\mathop{\mathrm{max}}\limits_{\pmb{\nu}\in\mathcal{D}} & \quad\quad \frac{\left|\pmb{\alpha}^H\pmb{\nu}+\theta\right|^2}{\left\|\pmb{A}\pmb{\nu}+\pmb{\gamma}\right\|^2+\varrho},
\end{align}
where $\mathcal{D}\triangleq \left\{\pmb{\nu} | \ |\nu_k|^2 \leq P_k ,\  k=1,2,\cdot\cdot\cdot,K\right\}$, $\pmb{\nu}\triangleq\left[\nu_1,\nu_2,\cdot\cdot\cdot,\nu_K\right]^T$, with $P_k$ being the power constraint of the $k^{\mathrm{th}}$ Eve, $\pmb{A}\triangleq\frac{1}{\sqrt{P_T}}\left[\pmb{h}_{E,1},\pmb{h}_{E,2},\cdot\cdot\cdot,\pmb{h}_{E,K}\right]$, $\pmb{\alpha}\triangleq \pmb{A}^H\pmb{h}_{E,K}$, and if $\pmb{h}_B$ is unknown, we have
$\{\varrho, \theta, \pmb{\gamma}  \}=\left\{\frac{P_S\sigma_{BT}^2}{\sigma_{E,K}^2}\left\|\pmb{h}_{E,K}\right\|^2 + N\sigma_{BT}^2,  0, \pmb 0 \right\}$, otherwise, we have $\{\varrho, \theta, \pmb{\gamma}  \}=\left\{\frac{P_S\sigma_T^2}{\tau P_T \sigma_{E,K}^2}\left\|\pmb{h}_{E,K}\right\|^2 + N\frac{\sigma_{T}^2}{\tau P_T}, \pmb{h}_{E,K}^H\pmb{h}_B, \pmb{h}_B \right\}$.
Note that  we assume an individual power constraint for each Eve in \eqref{TheSecondProblem} instead of a total power constraint, since the Eves are at different locations and can not share their power resources.

The objective function in \eqref{TheSecondProblem} is the ratio of two convex quadratic functions, which is generally not a concave function.
Therefore, problem \eqref{TheSecondProblem} is a non-convex optimization problem and difficult to solve.
Before providing an efficient algorithm to handle the problem, we first investigate the second scenario where the BS is more cautious and intelligent in ensuring secrecy transmissions.

\section{PSA design when the BS tries to detect the PSA}
In this section, we consider the case where the BS is cautious, and after each uplink training phase, performs
a detection operation based on the received pilot signal to determine whether a PSA has occurred.
Once the BS suspects that a PSA has been performed, it terminates the signal transmission phase to protect the confidential message.
Hence, the Eves have to be careful when performing the PSA to avoid being detected, i.e., they will try to conceal the PSA.
In  this section,  we assume that the Eves do not know the CSI between the BS and the LU, i.e., $\pmb{h}_B$. We first derive the probability that the PSA is successfully detected by the BS,
and then
we formulate the PSA design problem such that the SNR of the $K^{\mathrm{th}}$ Eve is maximized while  the probability of being detected is below a pre-designed limit.

\subsection{Successful Detection Probability}
The BS performs PSA detection before downlink data transmission.
The detection operation can be formulated as the following binary hypothesis test problem,
\begin{align}
\label{DetectionModel}
\pmb{y}_T=\left\{
\begin{aligned}
&\pmb{h}_B+\pmb{z}, && \mathcal{H}_0,\\
&\pmb{h}_B+\pmb{h}_E+\pmb{z}, && \mathcal{H}_1,
\end{aligned}\right.
\end{align}
where the null hypothesis $\mathcal{H}_0$ stands for the absence of the
PSA and the alternative hypothesis $\mathcal{H}_1$ represents
the presence of the PSA.
According to the detection model in \eqref{DetectionModel},
we derive expressions for the successful detection probability under the general case and the worst case for the Eves depending on the BS's prior knowledge of $\pmb{h}_E$.

\subsubsection{General case for the Eves}
In general, it is difficult for the BS to obtain the exact value of $\pmb{h}_E$. Therefore, the BS will model the aggregated channel $\pmb{h}_E$ for $\mathcal{H}_1$ in \eqref{DetectionModel} as an unknown complex-valued vector parameter. According to \eqref{DetectionModel}, the BS will model the distribution of $\pmb{y}_T$ as
\begin{align}\label{yTdistribution}
\pmb{y}_T\sim\left\{
\begin{aligned}
&\mathbb{CN}\left(\pmb{0},\sigma_{BT}^2\pmb{I}_N\right), && \mathcal{H}_0,\\
&\mathbb{CN}\left(\pmb{h}_E,\sigma_{BT}^2\pmb{I}_N\right), && \mathcal{H}_1,
\end{aligned}.\right.
\end{align}
Note that $\pmb{h}_E$ in \eqref{yTdistribution} is unknown by the BS, and therefore, we assume that it resorts to the generalized logarithm likelihood test \cite{BOOK:S.M.Kay} to distinguish between $\mathcal{H}_0$ and $\mathcal{H}_1$, which can be mathematically expressed as
\begin{align} T\left(\pmb{y}_T\right)&=\ln\frac{ \mathop{\mathrm{max}}\limits_{\pmb{h}_E} f\left(\pmb{y}_T|\mathcal{H}_1,\pmb{h}_E\right)}{f\left(\pmb{y}_T|\mathcal{H}_0\right)} = \frac{\left\|\pmb{y}_T\right\|^2}{\sigma_{BT}^2} \gtreqless_{\mathcal{H}_0}^{\mathcal{H}_1} \Lambda_G,
\label{logarithmlikelihoodGeneral}
\end{align}
where $\Lambda_G$ is the decision threshold which is set by the BS according to a pre-defined acceptable false alarm probability $\eta_G~\left(0<\eta_G<1\right)$.
The false alarm probability $\mathcal{P}_G^{(F)}$ is the probability that the BS mistakenly concludes that the PSA has occurred when the Eves are indeed not attacking, and can be expressed as
\begin{align}
\mathcal{P}_G^{(F)} \triangleq \mathbb{P}\left\{T\left(\pmb{y}_T\right)>\Lambda_G |\mathcal{H}_0\right\}.
\label{FalseAlarm}
\end{align}
The BS requires that the false alarm probability $\mathcal{P}_G^{(F)}$ does not exceed a pre-defined value $\eta$, i.e.,
\begin{align}
	\mathcal{P}_G^{(F)} &\overset{(a)}{=}
    \mathbb{P}\left\{ \left\|\pmb{y}_T\right\|^2 \geq \sigma_{BT}^2\Lambda_G|\mathcal{H}_0\right\} \nonumber \\
    &\overset{(b)}{=}\frac{1}{\Gamma_N}\int_{\sigma_{BT}^2\Lambda_G}
	^{+\infty}\frac{r^{N-1}}{\left(\sigma_{BT}^2\right)^N}\mathrm{e}^{-\frac{r}{\sigma_{BT}^2}}\mathrm{d}r \nonumber \\
	 &=\frac{1}{\Gamma_N}\Gamma_N\left(\Lambda_G\right) \leq \eta_G,
    \label{TheFalseAlarmProb}
\end{align}
where $(a)$ is obtained by inserting \eqref{logarithmlikelihoodGeneral} into \eqref{FalseAlarm} and $(b)$ follows from the fact that conditioned on $\mathcal{H}_0$, $\left\|\pmb{y}_T\right\|^2$ is a Gamma random variable with shape and scale parameters given by $N$ and $\sigma_{BT}^2$, respectively.
From \eqref{logarithmlikelihoodGeneral}, we observe that a higher value of $\Lambda_G$ results in a lower successful detection probability, and from \eqref{TheFalseAlarmProb}, we can see that $\mathcal{P}_G^{(F)}$ is monotonically decreasing in $\Lambda_G$. Therefore, we assume the BS sets $\mathcal{P}_G^{(F)}=\eta_G$ to maximize the successful detection probability, which leads to $\Lambda_G =\Gamma_N^{-1}\left(\eta_G\Gamma_N\right)$,
where $\Gamma_N^{-1}\left(y\right)$ is the inverse function of $\Gamma_N\left(x\right)$.
Inserting $\Lambda_G$ into \eqref{logarithmlikelihoodGeneral} leads to
\begin{align}
	\left\|\pmb{y}_T\right\|^2 \gtreqless_{\mathcal{H}_0}^{\mathcal{H}_1}E_G, \quad E_G \triangleq\sigma_{BT}^2\Gamma_N^{-1}\left(\eta_G\Gamma_N\right).
    \label{FinalDetector}
\end{align}
According to \eqref{FinalDetector},  the BS computes $\left\|\pmb{y}_T\right\|^2$ after each uplink channel training phase. If $\left\|\pmb{y}_T\right\|^2 < E_G$, it concludes that there is no PSA, otherwise, it assumes that the PSA has occurred.

On the other hand, the Eves know the value of the aggregate channel $\pmb h_E$. Therefore when attacking, they model the probability density function (PDF) of $\pmb{y}_T$ as
$f_E\left(\pmb{y}_T\right) = \frac{\mathrm{e}^{-\left\|\pmb{y}_T-\pmb{h}_E\right\|^2/\sigma_{BT}^2}}{\pi^N\left(\sigma_{BT}^2\right)^N}$.
Combining $f_E\left(\pmb{y}_T\right)$ with the detection criterion of the BS in \eqref{FinalDetector}, we obtain the successful detection probability of the PSA as a function of $\left\|\pmb{h}_E\right\|$, i.e.,
\begin{align}
&\mathcal{P}_G^{(D)}\left(\left\|\pmb{h}_E\right\|\right)\nonumber \\
\triangleq&\mathbb{P}\left\{\left\|\pmb{y}_T\right\|^2
>E_G|\mathcal{H}_1\right\}=\mathbb{P}\left\{\frac{2\left\|\pmb{y}_T\right\|^2}{\sigma_{BT}^2}
>2\Gamma_N^{-1}\left(\eta_G\Gamma_N\right)\right\} \nonumber \\
=&1-\mathrm{e}^{-\frac{\left\|\pmb{h}_E\right\|^2}{\sigma_{BT}^2}}\sum_{j=0}^{+\infty}
\frac{\left\|\pmb{h}_E\right\|^{2j}Q_{2N+2j}\left(2\Gamma^{-1}_N\left(\eta_G\Gamma_N\right)\right)}
{j!\sigma_{BT}^{2j}},
\label{DetectedProbWorstGeneral}
\end{align}
where $Q_k(x)$ is the cumulative distribution function (CDF) of a central chi-square distributed random variable with $k$ degrees of freedom defined as
$Q_k\left(x\right) = \frac{\gamma_{k/2}\left(\frac{x}{2}\right)}{\Gamma_{k/2}}$.

\subsubsection{Worst case for the Eves}
We also consider a worst case for the Eves where the aggregate channel $\pmb{h}_{E}$ is known by the BS during the detection procedure. In this case, from the perspective of the BS, the distribution of $\pmb{y}_T$ is given by
\begin{align}
	\pmb{y}_T\sim\left\{
	\begin{aligned}
		&\mathbb{CN}\left(\pmb{0},\sigma_{BT}^2\pmb{I}_N\right),& & \mathcal{H}_0,\\
		&\mathbb{CN}\left(\pmb{h}_{E},\sigma_{BT}^2\pmb{I}_N\right),& & \mathcal{H}_1,
	\end{aligned}\right.
\label{WorstCaseYtDistribution}
\end{align}
where $\pmb{h}_{E}$ now is a known parameter for the BS.
Assume the BS determines whether the PSA has occurred through the logarithm likelihood test.
Then, the logarithm likelihood ratio is given by
\begin{align}
\label{WorstCaseLikelihoddR}
	T\left(\pmb{y}_T\right)=\ln \frac{f\left(\pmb{y}_T|H_1\right)}{f\left(\pmb{y}_T|H_0\right)}&=\frac{\pmb{y}_T^H\pmb{h}_E+\pmb{h}_E^H\pmb{y}_T-\pmb{h}_E^H\pmb{h}_E}{\sigma_{BT}^2}.
\end{align}
By inserting \eqref{WorstCaseYtDistribution} into \eqref{WorstCaseLikelihoddR}, it can be proved that if $\mathcal{H}_0$ is true, then $T\left(\pmb{y}_T\right)$ is distributed as $\mathbb{N}\left(-\frac{\left\|\pmb{h}_E\right\|^2}{\sigma_{BT}^2}, \frac{2\left\|\pmb{h}_E\right\|^2}{\sigma_{BT}^2}\right)$, otherwise, the distribution of $T\left(\pmb{y}_T\right)$ is $\mathbb{N}\left(\frac{\left\|\pmb{h}_E\right\|^2}{\sigma_{BT}^2}, \frac{2\left\|\pmb{h}_E\right\|^2}{\sigma_{BT}^2}\right)$.
Thus, the decision between $\mathcal{H}_0$ and $\mathcal{H}_1$ corresponds to a decision between two Gaussian random variables with equal variance and opposite means.
The detection operation can be written as
\begin{align}
	T\left(\pmb{y}_T\right) \gtreqless_{\mathcal{H}_0}^{\mathcal{H}_1} \Lambda_W \label{TheDecisionCriterion},
\end{align}
where the BS set the threshold $\Lambda_W$ to ensure that the false alarm probability $\mathcal{P}_W^{(F)}$ is below an acceptable value $\eta_W$, i.e.,
\begin{align}
\mathcal{P}_W^{(F)} &= \mathbb{P}\left\{ T\left(\pmb{y}_T\right) > \Lambda_W | \mathcal{H}_0 \right\}\nonumber \\
&= \frac{\sigma_{BT}}{2\sqrt{\pi}\left\|\pmb{h}_{E}\right\|}\int_{\Lambda_W}^{+\infty}
\mathrm{e}^{ -\frac{\left( t + \left\|\pmb{h}_E\right\|^2/\sigma_{BT}^2\right)^2}{4\left\|\pmb{h}_E\right\|^2/\sigma_{BT}^2}} \mathrm{d}t \nonumber \\
&=\frac{1}{2}\left(1 - \Phi\left(\frac{\left\|\pmb{h}_E\right\|}{2\sigma_{BT}}+
\frac{\sigma_{BT}\Lambda_W}{2\left\|\pmb{h}_E\right\|}\right)\right)\leq \eta_W.
\end{align}
By setting $\mathcal{P}_W^{(F)} = \eta_W$, the BS obtains the maximal successful detection probability, and in this case, we have
\begin{align}
\label{WorstThreshold}
\Lambda_W = \frac{\left\|\pmb{h}_E\right\|}{\sigma_{BT}}
\left(2\Phi^{-1}\left(1 - 2\eta_W\right) - \frac{\left\|\pmb{h}_E\right\|}{\sigma_{BT}}\right),
\end{align}
where $\Phi^{-1}\left(\cdot\right)$ is the inverse function of the error function $\Phi\left(\cdot\right)$.
According to the decision criterion in \eqref{TheDecisionCriterion}, the successful detection probability is given by
\begin{align}
	&\mathcal{P}_W^{(D)}\left(\left\|\pmb{h}_E\right\|\right) = \mathbb{P}\left\{T\left(\pmb{y}_T\right) > \Lambda_W | \mathcal{H}_1 \right\} \nonumber \\
& =
    \frac{\sigma_{BT}}{2\sqrt{\pi}\left\|\pmb{h}_E\right\|}
	\int_{\Lambda_W}^{+\infty} \mathrm{e}^{-\frac{\left(t- \left\|\pmb{h}_E\right\|^2/\sigma_{BT}^2\right)^2}
{4 \left\|\pmb{h}_E\right\|^2/\sigma_{BT}^2 } } \mathrm{d}x \nonumber\\
&=
\frac{1}{2}\left(1 - \Phi\left(-\frac{\left\|\pmb{h}_E\right\|}{2\sigma_{BT}}+
\frac{\sigma_{BT}\Lambda_W}{2\left\|\pmb{h}_E\right\|}\right)\right)\nonumber \\
 &\overset{(*)}{=} \frac{1}{2}\left(1 - \Phi\left(-\frac{\left\|\pmb{h}_E\right\|}{\sigma_{BT}}+
\Phi^{-1}\left(1 - 2\eta_W\right)\right)\right),
\label{DetectedProbWorstCase}
\end{align}
where for step $(*)$, \eqref{WorstThreshold} was used.
So far, we have derived the probabilities that the PSA is successfully detected by the BS for the general and the worst cases in \eqref{DetectedProbWorstGeneral} and \eqref{DetectedProbWorstCase}, respectively. In the next subsection, we will use \eqref{DetectedProbWorstGeneral} and \eqref{DetectedProbWorstCase} to formulate the PSA design problem.

\subsection{Attacking Scheme Design}
If the Eves are aware that the BS will perform PSA detection, they will attempt to conceal their attack. More specifically, when they attack, they will try to make the PSA detection probability as small as possible to
trigger the transmission of the confidential signal in the downlink data transmission phase.
To achieve this, the Eves design the PSA signal such that the successful detection probability remains under an acceptable threshold $\epsilon$, i.e.,
\begin{align}
\mathcal{P}^{(D)}\left(\left\|\pmb{h}_E\right\|\right) \leq \epsilon,~~~0<\epsilon<1,
\label{OriginalProbConstraint}
\end{align}
where we have
\begin{align}
\mathcal{P}^{(D)}\left(\|\pmb{h}_E\|\right)\triangleq  \left\{
\begin{aligned}
&\mathcal{P}_G^{(D)}\left(\|\pmb{h}_E\|\right),&&\text{for general case},\\
&\mathcal{P}_W^{(D)}\left(\|\pmb{h}_E\|\right), &&\text{for worst case}.
\end{aligned}\right.
\end{align}
Based on \eqref{OriginalProbConstraint}, to maximize the SNR of the $K^{\mathrm{th}}$ Eve while avoiding being successfully detected, we establish the following optimization problem
\begin{subequations}
\label{CounterDetection}
\begin{align}
\mathop{\mathrm{max}}_{\pmb{\nu}\in\mathcal{D}} &\quad \frac{\left|\pmb{\alpha}^H\pmb{\nu}\right|^2}{\left\|\pmb{A}\pmb{\nu}\right\|^2+\varrho},\label{CounterDetectionObject} \\
\mathrm{s.t.} &\quad \mathcal{P}^{(D)}\left(\left\|\pmb{h}_E\right\|\right) \leq \epsilon.\label{CounterDetectionC2}
\end{align}
\end{subequations}
Compared to (\ref{TheSecondProblem}), problem \eqref{CounterDetection} is even more difficult to solve because of constraint \eqref{CounterDetectionC2}.
Fortunately, by checking the first order derivatives of both $\mathcal{P}_G^{(D)}\left(x\right)$ and $\mathcal{P}_W^{(D)}\left(x\right)$, we conclude that both $\mathcal{P}_G^{(D)}\left(x\right)$ and $\mathcal{P}_W^{(D)}\left(x\right)$ are monotonically increasing functions w.r.t. $x$ in the region $(0,+\infty)$. Therefore, we can  express constraint \eqref{CounterDetectionC2} equivalently as
\begin{align}
\|\pmb{h}_E\|^2=\|\pmb{A\nu}\|^2\leq \varpi^2,\label{PROEQ}
\end{align}
where $\varpi$ satisfies $\mathcal{P}^{(D)}\left(\varpi\right)=\epsilon$ and can be obtained by the bisection method due to the monotonicity of $\mathcal{P}^{(D)}(x)$. We note that \eqref{PROEQ} is a convex quadratic constraint.
Considering \eqref{PROEQ}, \eqref{CounterDetection} can be transformed into the following equivalent problem,
\begin{align}
\mathop{\mathrm{max}}\limits_{\pmb{\nu}\in\mathcal{D}} \quad \frac{\left|\pmb{\alpha}^H\pmb{\nu}\right|^2}{\left\|\pmb{A}\pmb{\nu}\right\|^2+\varrho},\quad\quad \mathrm{s.t.} \quad \|\pmb{A\nu}\|^2\leq \varpi^2. \label{CounterDetection2}
\end{align}
However, problem \eqref{CounterDetection2} is still  non-convex due to the non-convex objective function, and thus still difficult to solve.

In Section III-B and Section IV-B, we have formulated two non-convex optimization problems for the PSA, i.e., \eqref{TheSecondProblem} and \eqref{CounterDetection2}. In the next section, we propose an efficient algorithm to solve these problems.

\section{ MM-ADMM-based method for PSA optimization }
In this section, we present an efficient method to solve \eqref{TheSecondProblem} and \eqref{CounterDetection2}. We observe that \eqref{TheSecondProblem} and \eqref{CounterDetection2} are special cases of the following optimization problem
\begin{align}
\mathop{\mathrm{max}}\limits_{\pmb{\nu}\in\mathcal{D}} &\  S\left(\pmb{\nu}\right)\triangleq\frac{\left|\pmb{\alpha}^H\pmb{\nu}+\theta\right|^2}
{\left\|\pmb{A}\pmb{\nu}+\pmb{\gamma}\right\|^2+\varrho},\quad\mathrm{s.t.}\  \|\pmb{A\nu}\|^2\leq \varpi^2. \label{FinalProblem}
\end{align}
In fact, by setting $\varpi\rightarrow +\infty$, the constraint $\|\pmb{A\nu}\|^2\leq \varpi^2$ can be ignored and \eqref{FinalProblem} degrades to \eqref{TheSecondProblem}, and by setting $\theta=0$ and $\pmb{\gamma}=\pmb{0}$, \eqref{FinalProblem} simplifies to \eqref{CounterDetection2}.
Therefore, in this section, we focus on solving \eqref{FinalProblem}.

We have already established that the difficulty in solving \eqref{FinalProblem} lies in the non-convexity of the objective function.
The idea behind the proposed algorithm is to first transform the original non-convex problem into a series of convex problems by invoking the MM principle \cite{MMmethod1,MMmethod2} and solving them iteratively. In each iteration, one convex problem has to be solved.
Though general convex optimization tools, such as CVX \cite{CVX}, can solve the obtained convex problems,
these tools may not be efficient because they are designed for general convex programs.
Therefore, we further propose an ADMM-based low complexity algorithm to solve the obtained convex problem.
The ADMM algorithm decomposes the convex problem into several sub-problems, each of which either has a closed-form solution or can be solved by the bisection method or Newton's method.
Besides, part of the obtained sub-problems can be solved in parallel, which can be exploited for a more efficient implementation.
The whole MM-ADMM algorithm converges to a stationary point of the original non-convex problem.
We show via simulation that the resulting MM-ADMM method is computationally very efficient.
Brief introductions to the MM method and the ADMM algorithm are provided in Appendix \ref{MMIntroduction} and Appendix \ref{ADMMMMIntroduction}, respectively.

Finally, we provide an alternative method for solving a special case of \eqref{FinalProblem} when $\theta=0$ and $\pmb{\gamma}=\pmb{0}$ based on SDR. We also provide a sufficient condition for when the SDR-based method can achieve the global optimum. The SDR-based method will be used as a benchmark to verify the efficiency of the proposed MM-ADMM algorithm in Section VI.

\subsection{MM-ADMM-based Low Complexity Algorithm}
In the first step of our approach to solve \eqref{FinalProblem}, we use the MM method to transform the problem into a convex one.
The MM method maximizes the original objective function by iteratively maximizing a series of lower bounds on the original objective function. In each iteration, it constructs a lower bound to be employed as the current objective function based on the solution obtained in the previous iteration.
Assume the solution of problem \eqref{FinalProblem} obtained from the previous iteration is denoted by $\hat{\pmb{\nu}}$, then a lower bound on the original objective is provided in the following Lemma.
\begin{lemma}
\label{LowerBound}
A lower bound on the objective function of \eqref{FinalProblem} is given by
\begin{align}
\hat{S}\left(\pmb{\nu};\hat{\pmb{\nu}}\right)
&\triangleq
-a \left\|\pmb{A}\pmb{\nu} +\pmb{\gamma}\right\|^2
+ 2b\Re\left\{\pmb{\beta}^H\pmb{\nu}\right\}
- c\nonumber \\
&\leq \frac{\left|\pmb{\alpha}^H\pmb{\nu}+\theta\right|^2}{\left\|\pmb{A}\pmb{\nu}+\pmb{\gamma}\right\|^2+\varrho}
\end{align}
where $\pmb{\nu}\in \mathbb{C}^{K\times 1}$ is an arbitrary complex vector,
$a\triangleq\frac{\left|\pmb{\alpha}^H\hat{\pmb{\nu}}+\theta\right|^2}
		{\left(\left\|\pmb{A}\hat{\pmb{\nu}}+\pmb{\gamma}\right\|^2+\varrho\right)^2}$,
$b\triangleq\frac{1}{\left\|\pmb{A}\hat{\pmb{\nu}}+\pmb{\gamma}\right\|^2+\varrho}$,
$c\triangleq a\varrho+\frac{2\Re\left\{\left(\hat{\pmb{\nu}}^H\pmb{\alpha}+\theta^*\right)\theta\right\}}{\left\|\pmb{A}\hat{\pmb{\nu}}
+\pmb{\gamma}\right\|^2+\varrho}$, and
$\pmb{\beta}\triangleq \pmb{\alpha}\left( \pmb{\alpha}^H\hat{\pmb{\nu}}+\theta\right)$.
\end{lemma}
\begin{IEEEproof}
For any $x\in \mathbb{C}$ and $y\in (0,+\infty)$, $f\left(x;y\right)\triangleq\frac{\left|x\right|^2}{y}$ is jointly convex w.r.t. $(x,y)$ \cite{Book:S.Boyd2004}.
Therefore, $f\left(x;y\right)$ satisfies
$f\left(x;y\right)\geq f\left(\hat{x};\hat{y}\right) + 2\Re\left\{\left(x-\hat{x}\right)^*\nabla_{x^*}f\left(\hat{x};\hat{y}\right)\right\}+ \nabla_{y}f\left(\hat{x};\hat{y}\right)\left(y-\hat{y}\right)$,
where $\hat{x}\in \mathbb{C}$, $\hat{y}\in (0,+\infty)$, $\nabla_{x^*}f\left(\hat{x};\hat{y}\right)=\frac{\hat{x}}{\hat{y}}$, and $\nabla_{y}f\left(\hat{x};\hat{y}\right)=-\frac{\hat{x}^H\hat{x}}{\hat{y}^2}$.
By replacing $x$ with $\pmb{\alpha}^H\pmb{\nu}+\theta$ and $y$ with $\left\|\pmb{A}\pmb{\nu}+\pmb{\gamma}\right\|^2+\varrho$, we directly obtain Lemma 1.
\end{IEEEproof}
It can be verified that the lower bound provided in Lemma \ref{LowerBound} satisfies the conditions stated in Appendix \ref{MMIntroduction} for applying the MM method. Therefore, we can solve \eqref{FinalProblem} by iteratively solving the following problem
\begin{align}
	\mathop{\mathrm{max}}\limits_{\pmb{\nu}\in\mathcal{D}}  \quad \hat{S}\left(\pmb{\nu};\hat{\pmb{\nu}}\right), \quad\quad \mathrm{s.t.} \quad\|\pmb{A\nu}\|^2\leq \varpi^2.	\label{TheSecondProblemMMmethod}
\end{align}
Problem \eqref{TheSecondProblemMMmethod} maximizes a concave quadratic function under $K+1$ convex quadratic constraints, and thus is a convex problem.

Next, we propose an ADMM-based algorithm to obtain the global optimum of this convex problem.
The main advantage of the proposed method is that we can decompose problem \eqref{TheSecondProblemMMmethod} into several subproblems and each of them either has a closed-form solution or can be efficiently solved by the bisection method or Newton's method.
To this end, we introduce a new variable $\pmb{\Xi}\triangleq \left[\Xi_1,\Xi_2,\cdot\cdot\cdot,\Xi_K\right]^T$ and transform problem \eqref{TheSecondProblemMMmethod} into the following equivalent form,
\begin{align}
\label{TheSecondProblemMMmethodADMMFirstStep}
\mathop{\mathrm{min}}\limits_{\pmb{\nu}\in \mathcal{X}_1; \pmb{\Xi}\in\mathcal{X}_2} ~ a\left\|\pmb{A}\pmb{\nu}+\pmb{\gamma}\right\|^2 - 2b\Re\left\{\pmb{1}^T \pmb{\Xi}\right\}, ~\mathrm{s.t.}~ \pmb{\Xi} = \pmb{B}\pmb{\nu},
\end{align}
where $\mathcal{X}_2\triangleq\left\{\pmb{\Xi}|\left|\Xi_k\right|^2 \leq \left|\beta_k\right|^2P_k,\forall k=1,2,\cdot\cdot\cdot,K\right\}$, $\mathcal{X}_1\triangleq\left\{\pmb{\nu}|\left\|\pmb{A\nu}\right\|^2 \leq \varpi^2\right\}$, $\pmb{B}\triangleq \mathrm{diag}\left(\pmb{\beta}^*\right)$, $\pmb{1}\triangleq\left[1,1,\cdot\cdot\cdot,1\right]^T\in \mathbb{R}^{K\times 1}$, and $\beta_k$ is the $k^{\mathrm{th}}$ element of $\pmb{\beta}$.
According to the ADMM principle introduced in Appendix \ref{ADMMMMIntroduction},
the procedure for solving \eqref{TheSecondProblemMMmethodADMMFirstStep} consists of iterating the following updates from the $(n-1)^{\mathrm{th}}$ step to the $n^{\mathrm{th}}$ step,
\begin{subequations}
\label{ADMMIteration}
\begin{align}
\pmb{\nu}^{(n)} &= \mathop{\mathrm{argmin}}\limits_{\left\|\pmb{A\nu}\right\|^2 \leq \varpi^2}
\left\{\begin{aligned}
&a\left\|\pmb{A}\pmb{\nu}+\pmb{\gamma}\right\|^2+\frac{\rho}{2}\left\|\pmb{B}\pmb{\nu} - \pmb{\Xi}^{(n-1)}\right\|^2 \\
&+\Re\left\{\left\langle \pmb{y}^{(n-1)}, \pmb{B}\pmb{\nu} - \pmb{\Xi}^{(n-1)}\right\rangle\right\}
\end{aligned}\right\},
\label{ADMMIterationUpdateNu}\\
\pmb{\Xi}^{(n)} &= \mathop{\mathrm{argmin}}\limits_{\substack{ \left|\Xi_k\right|^2\leq\left|\beta_k\right|^2P_k  \\k=1,2,\cdot\cdot\cdot,K}}
\left\{\begin{aligned}
&- 2b\Re\left\{\pmb{1}^T \pmb{\Xi}\right\}+\frac{\rho}{2}\left\|\pmb{B}\pmb{\nu}^{(n)} - \pmb{\Xi}\right\|^2\\
&+\Re\left\{\left\langle \pmb{y}^{\left(n-1\right)},\pmb{B}\pmb{\nu}^{(n)} - \pmb{\Xi} \right\rangle\right\}
\end{aligned}\right\},
\label{ADMMIterationUpdateXi}\\
\pmb{y}^{(n)} &= \pmb{y}^{(n-1)} + \rho\left(\pmb{B}\pmb{\nu}^{(n)} - \pmb{\Xi}^{(n)}\right),
\end{align}
\end{subequations}
where  $\rho$ can be chosen as any positive real number and $\{\pmb{\nu}^{(n)}, \pmb{\Xi}^{(n)},\pmb{y}^{(n)}\}$ denotes the results obtained in the $n^{\mathrm{th}}$ iteration.
The iterations in \eqref{ADMMIteration} can be efficiently carried out as follows.

\subsubsection{Update of $\pmb{\nu}^{(n)}$}
In fact, \eqref{ADMMIterationUpdateNu} is a convex quadratic problem, which means the Karush-Kuhn-Tucker (KKT) conditions are sufficient and necessary for the globally optimal solution \cite{Book:S.Boyd2004}.
The KKT conditions for \eqref{ADMMIterationUpdateNu} are
\begin{subequations}
\begin{align}
&\left(a+\zeta\right)\pmb{T}\pmb{\nu} + \frac{\rho}{2}\pmb{Y}\pmb{\nu} -\pmb{\mu} = \pmb{0},\label{KKTNU1}\\
&~\zeta \geq 0,~\left\|\pmb{A\nu}\right\|^2 \leq \varpi^2,~\zeta\left(\left\|\pmb{A\nu}\right\|^2 - \varpi^2\right)=0,
\label{KKTNU2}
\end{align}
\end{subequations}
where $\pmb{T}\triangleq\pmb{A}^H\pmb{A}$, $\pmb{Y}\triangleq\pmb{B}^H\pmb{B}$, $\pmb{\mu}\triangleq\frac{\rho}{2}\pmb{B}^H\pmb{\Xi}^{(n-1)} - \frac{1}{2}\pmb{B}^H\pmb{y}^{(n-1)}$, and $\zeta$ is the dual variable w.r.t. the constraint $\left\|\pmb{A\nu}\right\|^2 \leq \varpi^2$. According to \eqref{KKTNU1}, $\pmb{\nu}$ is a function of $\zeta$ as
\begin{align}
\pmb{\nu}\left(\zeta\right)=\left(\left(a+\zeta\right)\pmb{T}+\frac{\rho}{2}\pmb{Y}\right)^{-1}\pmb{\mu}.
\label{KKTNU1-1}
\end{align}
According to \eqref{KKTNU2}, if $\|\pmb{A}\pmb{\nu}\left(0\right)\|^2\leq \varpi^2$, then we have $\pmb{\nu}^{(n)}=\pmb{\nu}\left(0\right)$,
otherwise, we have to search for $\zeta^\prime$ such that $\pmb{\nu}\left(\zeta^\prime\right)^H\pmb{T}\pmb{\nu}\left(\zeta^\prime\right)=0$ within $(0,+\infty)$.
Note that $\pmb{\nu}\left(\zeta\right)^H\pmb{T}\pmb{\nu}\left(\zeta\right)=\sum_{k=1}^K\frac{\mu_k^\prime\Pi_k}{(\rho/2+(a+\zeta)\Pi_k)^2}$ is a convex decreasing function w.r.t. $\zeta$, where $\pmb{\mu}^\prime~=~[\mu_1^\prime,\mu_2^\prime\cdots,\mu_K^\prime]^T~=~\pmb{Q}^H\pmb{Y}^{-\frac{1}{2}}\pmb{\mu}$ and $\pmb{Z}~=~\pmb{Q}\mathrm{diag}\left\{\pmb{\Pi}\right\}\pmb{Q}^H$ is the eigenvalue decomposition with $\pmb{Z}~\triangleq~\pmb{Y}^{-\frac{1}{2}}\pmb{T}\pmb{Y}^{-\frac{1}{2}}$ and $\pmb{\Pi}$, with its $k^{\mathrm{th}}$ element denoted by $\Pi_k$, is a vector  containing all the eigenvalues of $\pmb{Z}$. Therefore, we can solve $\pmb{\nu}\left(\zeta^\prime\right)^H\pmb{T}\pmb{\nu}\left(\zeta^\prime\right)=0$ using the bisection method or Newton's method \cite{K.Huang2016}. Once we obtain $\zeta^\prime$, we have $\pmb{\nu}^{(n)}=\pmb{\nu}\left(\zeta^\prime\right)$.

Note that for solving \eqref{TheSecondProblem}, the constraint $\left\|\pmb{A\nu}\right\|^2 \leq \varpi^2$ in \eqref{FinalProblem} is absent. Hence, \eqref{ADMMIterationUpdateNu} becomes an unconstraint convex quadratic problem, and we can obtain the optimal solution in closed form by the first order condition \cite{Book:S.Boyd2004}. Setting the first order derivative to zero, we directly obtain
\begin{align}
\pmb{\nu}^{(n)} = \left(a\pmb{T} + \frac{\rho}{2}\pmb{Y}\right)^{-1}\pmb{\mu}. \label{OptimalNuExpression}
\end{align}

\subsubsection{Update of $\pmb{\Xi}^{(n)}$}
Problem \eqref{ADMMIterationUpdateXi} contains $K$ inequality constraints, which complicates the problem. However, by exploiting its special structure, we can decompose \eqref{ADMMIterationUpdateXi} into $K$ parallel sub-problems, and the optimal solutions of these sub-problems is the optimal solution of \eqref{ADMMIterationUpdateXi}. The $k^{\mathrm{th}}$ sub-problem is
\begin{align}
\label{TheFirstProblemADMMSub1}
\mathop{\mathrm{min}}\limits_{ \left|\Xi_k\right|^2 \leq \left|\beta_k\right|^2P_k } \left\{
\begin{aligned}
&\frac{\rho}{2} \left|\beta_k^*\nu_k^{(n)} - \Xi_k\right|^2 -
2b\Re\left\{\Xi_k\right\} \\
&+ \Re\left\{\left(y_k^{(n-1)}\right)^*\left(\beta_k^*\nu_k^{(n)} - \Xi_k\right)\right\}
\end{aligned}\right\}.
\end{align}
In fact, this sub-problem has a closed-form solution.
The Lagrangian function of \eqref{TheFirstProblemADMMSub1} is given by
\begin{align}
L_k\left( \Xi_k, \lambda_k \right) &= \frac{\rho}{2} \left|\beta_k^*\nu_k^{(n)} - \Xi_k\right|^2- \Re\left\{2b\Xi_k\right\} \nonumber\\
&\quad +\Re\left\{\left(y_k^{(n-1)}\right)^*\left(\beta_k^*\nu_k^{(n)} - \Xi_k\right)\right\} \nonumber\\
&\quad +\lambda_k\left( \left|\Xi_k\right|^2 - \left|\beta_k\right|^2P_k \right),\nonumber
\end{align}
where $\lambda_k\geq0$ is the dual variable w.r.t. constraint $ \left|\Xi_k\right|^2 \leq \left|\beta_k\right|^2P_k$. The first order condition is
\begin{align}
	\frac{\partial L_k}{\partial \Xi_k^*} = \frac{\rho\left(\Xi_k-\beta_k^*\nu_k^{(n)}\right)}{2}-b-\frac{y_k^{(n-1)}}{2}+\lambda\Xi_k=0,
\end{align}
and therefore, we obtain
\begin{align}
	\Xi_k\left(\lambda_k\right) = \frac{b+\frac{y_k^{(n-1)}}{2} + \frac{\rho}{2}\beta_k^*\nu_k^{(n)}}{\frac{\rho}{2}+\lambda_k}.
\end{align}
According to the complementary slackness condition \cite{Book:S.Boyd2004}, the optimal dual variable $\lambda_k^{\mathrm{opt}}$ has to satisfy
\begin{align}
\begin{aligned}
&\left|\Xi_k\left(\lambda_k^{\mathrm{opt}}\right)\right|^2 \leq \left|\beta_k\right|^2P_k,\quad \lambda_k\geq 0,\\
&\lambda_k^{\mathrm{opt}} \left( \left|\Xi_k\left(\lambda_k^{\mathrm{opt}}\right)\right|^2 - \left|\beta_k\right|^2P_k \right) = 0.
\end{aligned}
\label{ComplementarySlackCondition}
\end{align}
According to \eqref{ComplementarySlackCondition}, if $\left|\Xi_k\left(0\right)\right|^2 \leq \left|\beta_k\right|^2P_k$, we have  $\lambda_k^{\mathrm{opt}}=0$ and
$\Xi_k^{\mathrm{opt}}=\Xi_k\left(0\right)$.
Otherwise, we have to solve the equation $\left|\Xi_k\left(\lambda_k^{\mathrm{opt}}\right)\right|^2 - \left|\beta_k\right|^2P_k=0$, which leads to
\begin{align}
\label{GammaOptimal}
\begin{aligned}
\lambda_k^{\mathrm{opt}}&=\frac{\left|b+\frac{y_k^{(n-1)}}{2} +
\frac{\rho}{2}\beta_k^*\nu_k^{(n)}\right|}{\left|\beta_k\right|\sqrt{P_k}}-\frac{\rho}{2},\\
\Xi_k^{\mathrm{opt}}&=\Xi_k\left(\lambda_k^{\mathrm{opt}}\right).
\end{aligned}
\end{align}

Combining all steps from \eqref{TheSecondProblemMMmethod} to \eqref{GammaOptimal}, we now summarize the  algorithm proposed for solving \eqref{FinalProblem} in Algorithm \ref{MMandADMM}, where $\delta_M$ and $\delta_A$ are small positive numbers determining the accuracy of the MM and the ADMM algorithms, respectively, and
$T_{M}^{MAX}$ and $T_{A}^{MAX}$ are the maximum numbers of iterations allowable for the MM and the ADMM algorithms, respectively.
\begin{algorithm}[t]
\caption{MM-ADMM-based algorithm for solving \eqref{FinalProblem}}
\label{MMandADMM}
\begin{algorithmic}[1]
\STATE Initialize $\hat{\pmb{\nu}}$ satisfying $\left|\hat{\nu}_k\right|^2\leq P_k,~k=1,2,\cdot\cdot\cdot,K$, and $\left\|\pmb{A}\hat{\pmb{\nu}}\right\|\leq\varpi^2$;	
\STATE Initialize $n_M = 0$;	
\STATE \textbf{Repeat:}
\STATE $~~~$$n_M = n_M + 1$;
\STATE $~~~$Compute $a,~b,~\pmb{\beta}$ according to Lemma 1 and set $\pmb{B}=\mathrm{diag\left(\pmb{\beta}^*\right)}$;
\STATE $~~~$Initialize $\pmb{\Xi}^{(0)}=\hat{\pmb{\nu}},~\pmb{y}^{(0)}=\pmb{0}$ and set $n_A=0$, $\mathrm{SNR}_{\mathrm{Ini}} = S\left(\hat{\pmb{\nu}}\right)$;
\STATE $~~~$\textbf{Repeat:}
\STATE $~~~~~~$$n_A~=~n_A+1$;
\STATE $~~~~~~$Compute $\pmb{\nu}^{(n_A)}$ according to \eqref{KKTNU1-1} or \eqref{OptimalNuExpression};
\STATE $~~~~~~$Compute $\pmb{\Xi}^{(n_A)}$ according to \eqref{GammaOptimal};
\STATE $~~~~~~$Update $\pmb{y}^{(n_A)} = \pmb{y}^{(n_A-1)} + \rho\left(\pmb{B}\pmb{\nu}^{(n_A)}-\pmb{\Xi}^{(n_A)}\right)$;
\STATE $~~~$\textbf{Until:} $\left|\frac{\hat{S}\left(\pmb{\nu}^{(n_A)};\hat{\pmb{\nu}}\right) -\hat{S}\left(\pmb{\nu}^{(n_A-1)};\hat{\pmb{\nu}}\right)}{\hat{S}\left(\pmb{\nu}^{(n_A)};\hat{\pmb{\nu}}\right)}\right| < \delta_A$ or $n_A \geq T_{A}^{MAX}$.
\STATE $~~~$Set $\hat{\pmb{\nu}}=\pmb{\nu}^{(n_A)}$;
\STATE \textbf{Until:} $\left|\frac{S\left(\hat{\pmb{\nu}}\right) - \mathrm{SNR}_{\mathrm{Ini}}}{S\left(\hat{\pmb{\nu}}\right)}\right|<\delta_M$ or $n_M\geq T_{M}^{MAX}$.
\end{algorithmic}
\end{algorithm}

\subsection{Initialization, Convergence, and Complexity Analysis of the MM-ADMM Algorithm}
\subsubsection{Initialization}
To run Algorithm \ref{MMandADMM}, we need to initialize $\hat{\pmb{\nu}}$, i.e., step 1 in Algorithm \ref{MMandADMM}.
This can be easily done by randomly generating a vector $\breve{\pmb{\nu}}$ and setting $\hat{\pmb{\nu}}=\chi\breve{\pmb{\nu}}$ where $\chi$ is a scaling factor which is chosen such that all constraints are fulfilled.

\subsubsection{Convergence}
We observe that
\begin{enumerate}[1)]
\item the ADMM algorithm in \eqref{ADMMIteration} is guaranteed to converge to the global optimal solution of \eqref{TheSecondProblemMMmethod}
\cite{S.Boyd2011,D.P.Bertisekas1989,K.Huang2016,E.ChenTCOM2017};
\item by iteratively solving \eqref{TheSecondProblemMMmethod}, the MM method will converge to a stationary point of the original problem \eqref{FinalProblem} \cite{MMmethod1,MMmethod2}.
\end{enumerate}
Based on these two observations, we conclude that the MM-ADMM algorithm converges to a stationary point of problem \eqref{FinalProblem}.

\subsubsection{Complexity analysis}
The MM-ADMM algorithm solves \eqref{FinalProblem} by iteratively solving \eqref{TheSecondProblemMMmethodADMMFirstStep} via the ADMM algorithm.
Therefore, we only analyze the complexity of solving \eqref{TheSecondProblemMMmethodADMMFirstStep}.
Using the ADMM algorithm, we iteratively update $\{\pmb{\nu},~\pmb{\Xi},~\pmb{y}\}$ in (38). Updating $\pmb{\nu}$ requires performing an eigenvalue decomposition w.r.t. matrix $\pmb{Z}$ and solving an equation via Newton's method.
The complexity of these operations is on the order of $K^3+J_N K$, where $J_N$ is the number of iterations for Newton's method.
The updates of $\pmb{\Xi}$ and $\pmb{y}$ only require multiplications, and the corresponding complexity is on the order of $K$.
We note that when iterating \eqref{ADMMIteration}, we only need to perform the eigenvalue decomposition once because matrix $\pmb{Z}$ remains constant during the whole iterative procedure.
Therefore, the complexity of solving \eqref{TheSecondProblemMMmethodADMMFirstStep} using the ADMM algorithm is on the order of  $K^3+J_AJ_NK$, where $J_A$ is the number of iterations needed for convergence. The numerical results in Section VI show that tens of iterations are generally enough for the algorithm to converge to a solution with a relative error less than $10^{-4}$.
Since \eqref{TheSecondProblemMMmethodADMMFirstStep} is a convex problem, it can also be solved by general convex optimization tools.
For example, the SDPT3 solver of CVX solves \eqref{TheSecondProblemMMmethodADMMFirstStep} by the interior method,
which requires the computation of the inverse of a Hessian matrix in each iteration.
This entails a complexity on the order of $K^3$.
The number of iterations required to reach an $\varepsilon$-optimal solution is on the order of $\sqrt{K}\ln\left(\frac{1}{\varepsilon}\right)$
\cite{A.Ben-Tal2001,K.Y.Wang2014}.
Therefore, the complexity of solving \eqref{TheSecondProblemMMmethodADMMFirstStep} using the SDPT3 solver is about $\ln\left(\frac{1}{\varepsilon}\right)\sqrt{K}\times K^{3}$.
If we set $\varepsilon = 10^{-4}$, the computational complexity is about $9.2\sqrt{K}\times K^{3}$.
Compared with the ADMM algorithm, using CVX for solving \eqref{TheSecondProblemMMmethodADMMFirstStep} is more complex and also more time consuming as will be shown in the simulation section.

\subsection{An SDR-based Method}
In this subsection, we provide an alternative SDR-based method to solve a special case of \eqref{FinalProblem}, namely, when the Eves do not know the value of $\pmb{h}_B$. In this case, we have $\theta=0$ and $\pmb{\gamma}=\pmb{0}$.
This method is used as a benchmark for the MM-ADMM algorithm developed in the previous subsection.
First, defining $\pmb{V}=\pmb{\nu}\pmb{\nu}^H$, we rewrite \eqref{FinalProblem} in the following equivalent form,
\begin{subequations}
\label{CounterDetectionSDP1}
\begin{align}
\mathop{\mathrm{max}}_{\pmb{V}\succeq\pmb{0}} &\quad
\frac{\mathrm{Tr}\left(\pmb{\Theta}\pmb{V}\right)}{\mathrm{Tr}\left(\pmb{T}\pmb{V}\right)+\varrho},\label{CounterDetectionSDP1Object} \\
\mathrm{s.t.}  &\quad \mathrm{Tr}\left(\pmb{D}_k\pmb{V}\right) \leq P_k, \quad \forall k=1,2,\cdot\cdot\cdot,K, \label{CounterDetectionSDP1C1} \\
&\quad \mathrm{Tr}\left(\pmb{T}\pmb{V}\right)\leq \varpi^2, \label{CounterDetectionSDP1C2} \quad ~\mathrm{rank}\left(\pmb{V}\right)=1,
\end{align}
\end{subequations}
where $\pmb{T}\triangleq \pmb{A}^H\pmb{A}$, $\pmb{\Theta}\triangleq\pmb{\alpha}\pmb{\alpha}^H$, and $\pmb{D}_k$ is a $K\times K$ matrix whose $k^{\mathrm{th}}$ entry on the main diagonal is $1$ and all other entries are $0$.
Then, we introduce two new variables $\pmb{X}$ and $\kappa$ and apply the Charnes-Cooper transformation \cite{A.Charnes1962} by setting $\pmb{V}=\pmb{X}/\kappa$. In this way, \eqref{CounterDetectionSDP1} becomes
\begin{subequations}
	\label{CounterDetectionSDP3}
	\begin{align}
		\mathop{\mathrm{max}}\limits_{\substack{\pmb{X}\succeq\pmb{0};\kappa\\ \mathrm{rank}\left(\pmb{X}\right)=1}} &\quad
        \mathrm{Tr}\left(\pmb{\Theta}\pmb{X}\right),\label{CounterDetectionSDP3Object} \\
		\mathrm{s.t.}\   &\quad \mathrm{Tr}\left(\pmb{D}_k\pmb{X}\right) - \kappa P_k \leq 0,\  \forall 1\leq k\leq K, \label{CounterDetectionSDP3C1}\\
		&\quad \mathrm{Tr}\left(\pmb{T}\pmb{X}\right)- \kappa\varpi^2\leq 0, \label{CounterDetectionSDP3C2}\\
        &\quad \mathrm{Tr}\left(\pmb{T}\pmb{X}\right)+\kappa\varrho  = 1,\label{CounterDetectionSDP3C3}
	\end{align}
\end{subequations}
By using the SDR technique, we ignore the non-convex \emph{rank-one} constraint and \eqref{CounterDetectionSDP3} becomes
	\begin{align}
		\mathop{\mathrm{max}}_{\pmb{X}\succeq\pmb{0};\kappa} &\quad
        \mathrm{Tr}\left(\pmb{\Theta}\pmb{X}\right),\quad
		\mathrm{s.t.}\quad \eqref{CounterDetectionSDP3C1}-\eqref{CounterDetectionSDP3C3}, \label{CounterDetectionSDP4Object}
	\end{align}
which is a convex semidefine program (SDP) and can be solved by general convex optimization tools.
In general, \eqref{CounterDetectionSDP4Object} is not equivalent to \eqref{CounterDetectionSDP3} due to the omission of the \emph{rank-one} constraint.
However, if the optimal solution of \eqref{CounterDetectionSDP4Object} is \emph{rank-one}, then it is also an optimal solution of \eqref{CounterDetectionSDP3}.
A sufficient condition for the optimal solution of \eqref{CounterDetectionSDP4Object} to be \emph{rank-one} is provided in the following proposition.
\begin{proposition}
\label{ppp}
If the number of the Eves, $K$, and the number of the antennas at the BS, $N$, satisfy $K\leq N$, then the solution to problem \eqref{CounterDetectionSDP4Object} is a rank-one solution with probability one.
\end{proposition}
\begin{IEEEproof}
Please refer to Appendix \ref{ProofP}.
\end{IEEEproof}
{ In practice, both scenarios with $K\leq N$ and $K>N$ are possible.
If $K\leq N$, according to Proposition 1, the SDR-based method provides a global optimal solution of \eqref{CounterDetectionSDP3}. If $K>N$, a rank-one solution can not be guaranteed, and thus the SDR-based solution provides only an upper bound for \eqref{CounterDetectionSDP3} due to the relaxation of the rank constraint. However,
exhaustive numerical experiments, see Section VI, suggest that
the solution obtained with the SDR-based method is rank-one or near rank-one even when $K>N$, and thus near globally optimal.
The results obtained with the SDR-based method will serve as a benchmark to evaluate the efficiency of the MM-ADMM algorithm in Section VI.}

\emph{Complexity analysis and comparison:} In \eqref{CounterDetectionSDP4Object}, there are $K^2$ variables, $K+2$ inequality constraints, and one equality constraint.
Therefore, the complexity of solving \eqref{CounterDetectionSDP4Object} is on the order of  $K^{6.5}$ \cite{A.Ben-Tal2001,K.Y.Wang2014}.
Compared to the complexity of the MM-ADMM algorithm in Algorithm \ref{MMandADMM}, the SDR-based method is considerably more complex, which will be confirmed by simulations in the next section.

\section{Numerical Results}
In this section, we provide numerical results to illustrate the efficiency of the proposed method and the impact of the PSA performed by multiple Eves.
In our simulations, we normalize $\sigma_T^2=\sigma_{E,k}^2=0~\mathrm{dBm}, k=1,2,\cdot\cdot\cdot, K$, for convenience and we
assume that all Eves have the same attacking power, i.e., $P_k=P, k=1,2,\cdot\cdot\cdot, K$.
To terminate the MM-ADMM algorithm given in Algorithm \ref{MMandADMM}, we set $T_A^{MAX}=5$, $T_M^{MAX}=500$, $\delta_M = 10^{-3}$, and $\delta_M = 10^{-4}$ in all simulations.

\subsection{Convergence of the MM-ADMM Algorithm}
\begin{figure}
    \centering
  \subfigure[SNR versus the iteration steps.]{
    \label{fig:subfig:a} 
    \includegraphics[width=2.7 in ]{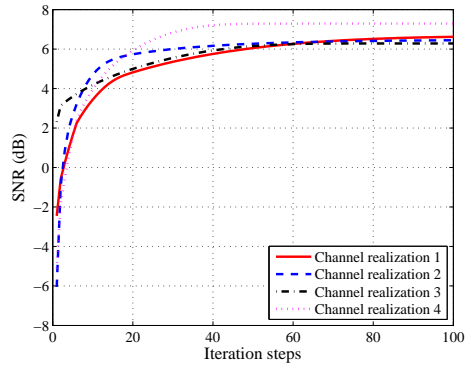}}
  \subfigure[Number of iteration steps required for convergence versus random channel realization index.]{
    \label{fig:subfig:b} 
    \includegraphics[width=2.7 in ]{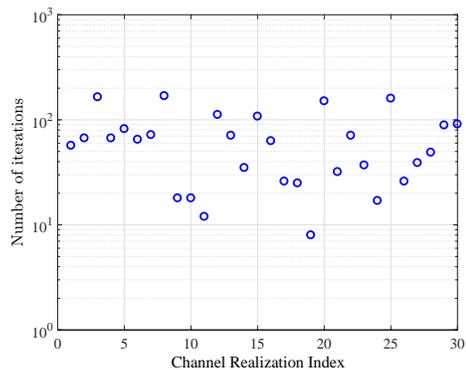}}
  \caption{The convergence of the MM-ADMM algorithm. We set $N=8$, $K=3$, $P_S=20$ (dBm), $P_T=10$ (dBm), $P=10$ (dBm), and $\rho = 0.01$.}
  \label{fig:subfig} 
  \vspace{-5 mm}
\end{figure}
The convergence behavior of Algorithm \ref{MMandADMM} is shown in Fig. \ref{fig:subfig}.
In Fig. \ref{fig:subfig:a}, we plot the optimized SNRs versus the iteration steps for $4$ random channel realizations.
From Fig. \ref{fig:subfig:a}, we observe that for all considered channel realizations,  the MM-ADMM algorithm converges.
In Fig. \ref{fig:subfig:b},
for 30 randomly generated channel realizations, the numbers of iterations needed for the algorithm to converge are plotted. As can be observed, 100 iterations are enough for the MM-ADMM  algorithm to converge in most cases.

\subsection{Efficiency of the Proposed MM-ADMM Algorithm}
\begin{figure*}
    \centering
  \subfigure[The comparison of the average simulation time of different methods.]{
    \label{RunningTimevsK} 
    \includegraphics[width=2 in ]{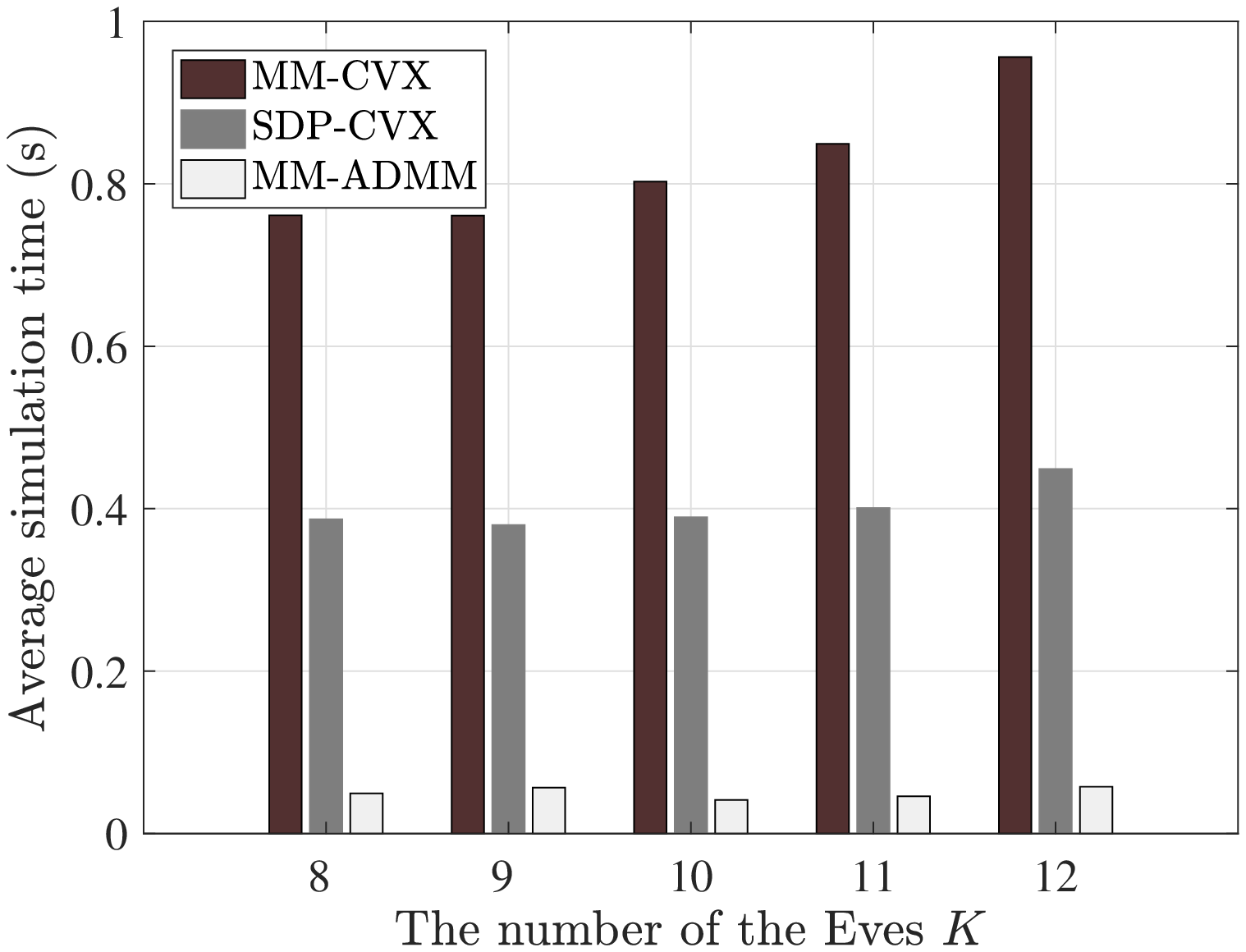}}
    \hspace{0.05 in}
  \subfigure[The eigenvalues of the solution of the SDR-based method when $K=13$.]{
    \label{SumAgainstDetectionNoHbEigenValueSDP} 
    \includegraphics[width=2 in ]{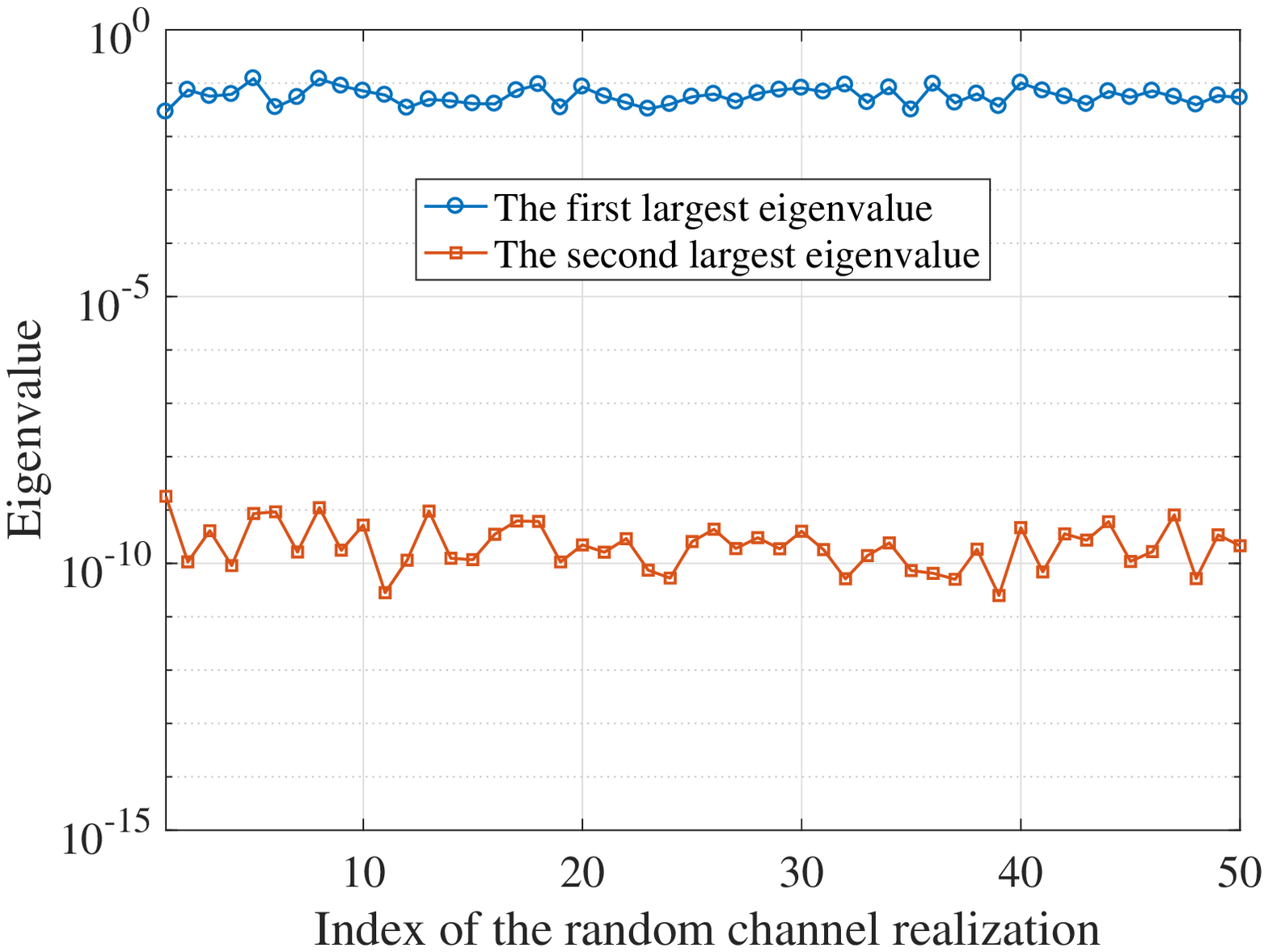}}
    \hspace{0.05 in}
  \subfigure[The wiretapping SNR versus the number of Eves.]{
    \label{SumAgainstDetectionNoHbSNRvsK} 
    \includegraphics[width=2 in ]{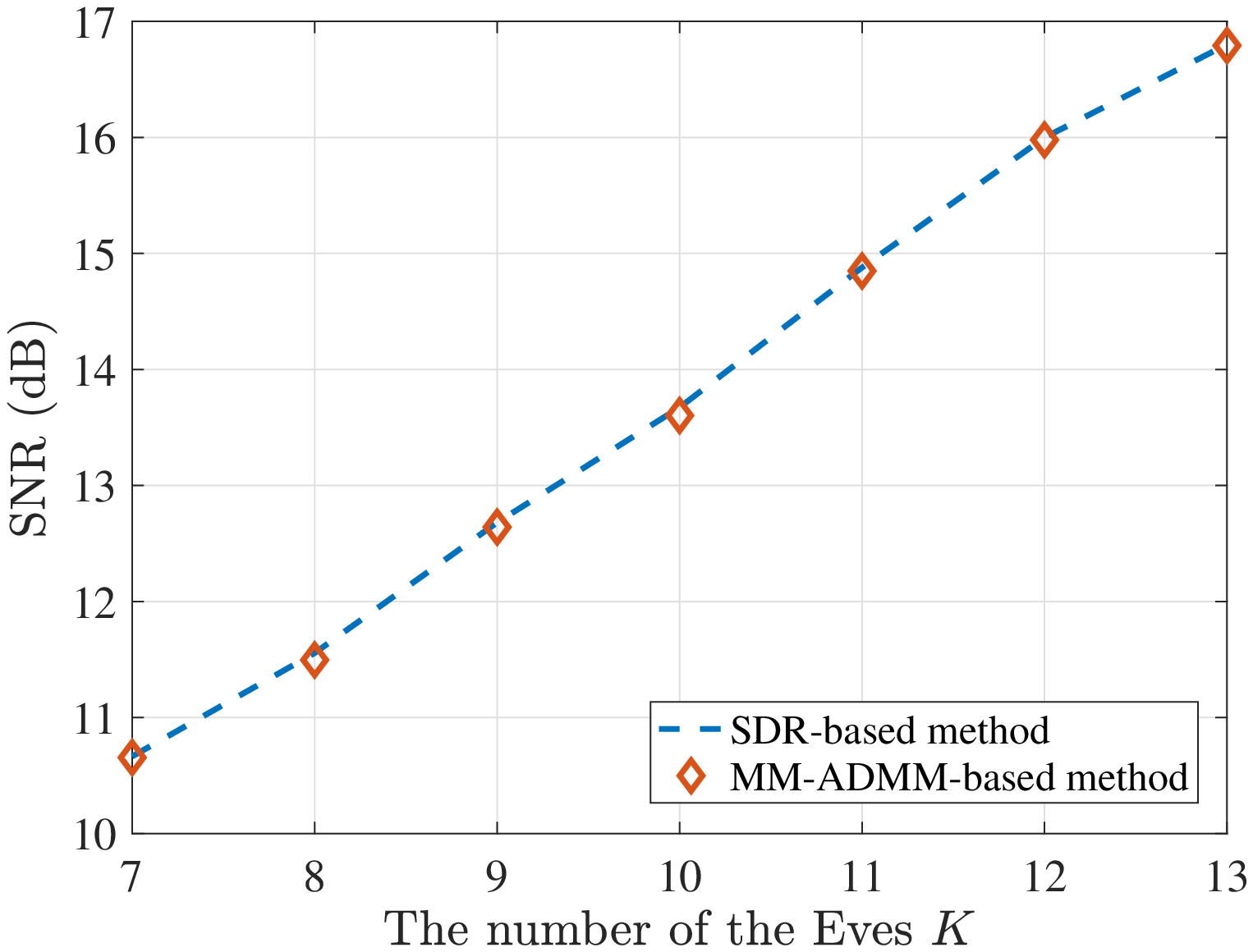}}
  \caption{Comparison of different methods, where $N=10$, $P=8$ (dBm), $P_T=10$ (dBm), $P_S=20$ (dBm), $\eta_G=0.05$, and $\epsilon=0.2$.}
  \label{EfficiencyOfMMADMM} 
  \hrulefill
  \vspace{-5 mm}
\end{figure*}

To show the efficiency of the MM-ADMM algorithm, we compare the MM-ADMM algorithm with other methods in terms of the average time \footnote{
The complexity of the MM-ADMM algorithm depends on the number of iterations required for convergence, which is difficult to analyze. Therefore, to be able to compare the relative computational complexities of different methods, we resort to the average simulation time as performance criterion. Note that using the average simulation time to compare the computational complexity of different optimization algorithm is common in the related literature, see e.g., \cite{AverageTime1,AverageTime2}.} required for obtaining the final solution and the wiretapping performance in Fig. \ref{EfficiencyOfMMADMM}.
In Fig. \ref{EfficiencyOfMMADMM}, we consider the case where the BS has no prior knowledge of $\pmb{h}_E$ and attempts to detect the PSA by setting the false alarm probability as $\eta_G = 0.05$.

In Fig. \ref{RunningTimevsK}, we compare the MM-ADMM algorithm in Algorithm \ref{MMandADMM} with the following two methods in terms of the time required to solve \eqref{FinalProblem},
\begin{enumerate}
    \item \emph{SDR-CVX:} This method is described in Section V-B, and is guaranteed to find the global optimum for $K\leq N$.

    \item \emph{MM-CVX:} In this method, we use the MM method to transform \eqref{FinalProblem} into \eqref{TheSecondProblemMMmethod}, and use CVX to solve \eqref{TheSecondProblemMMmethod}.
\end{enumerate}
The average simulation times in Fig. \ref{RunningTimevsK} are obtained by using the timing instruction of the commercial \textit{MATLAB} software, i.e., `tic' and `toc', and are averaged across 1000 random channel realizations.
As can be observed, the proposed MM-ADMM method is much faster than the two competing methods.

{In Fig. \ref{SumAgainstDetectionNoHbEigenValueSDP}, we investigate the rank of the solution obtained with the SDR-based method when $K>N$. To this end, we plot the largest and the second largest eigenvalues of the solutions $\pmb{X}$ for $50$ random channel realizations.
From Fig. \ref{SumAgainstDetectionNoHbEigenValueSDP}, we see that the second largest eigenvalues are orders of magnitude smaller than the largest eigenvalues, i.e., the obtained solutions are approximately \emph{rank-one}.
In fact, we have done a vast number of numerical experiments, but failed to obtain a solution where the ratio between the second largest and the largest eigenvalues exceeded $10^{-6}$.
This means that though without theoretical guarantee, the SDR-based method is expected to find a near-global optimum in many cases, even when $K>N$.}

{In Fig. \ref{SumAgainstDetectionNoHbSNRvsK}, we show the wiretapping SNR achieved by the MM-ADMM algorithm and the upper bound of the wiretapping SNR obtained with the SDR-based method (due to the relaxation of the rank-one constraint). The results are averaged over 1000 random channel realizations.
From Fig. \ref{SumAgainstDetectionNoHbSNRvsK}, we can see that the MM-ADMM method achieves almost the same performance as the upper bound obtained with the SDR-based method for both $K\leq N$ and $K>N$.}

In summary, the proposed MM-ADMM algorithm is computationally more efficient than the SDR-based method but achieves practically the same performance.

\subsection{The BS is Unaware of the PSA}
\begin{figure}
    \centering
  \subfigure[The wiretapping SNR versus the number of Eves $K$. We set $P_T=10$ (dBm) and $P_S=20$ (dBm).]{
    \label{SchemeCom} 
    \includegraphics[width=2.7 in ]{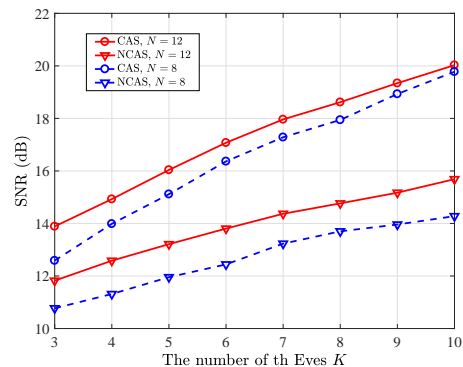}}
  \subfigure[The wiretapping SNR versus the training power of the LU. We set $N=12$, $P=5$ (dBm) and $P_S=10$ (dBm).]{
    \label{SumNoDetectionSNRvsPt} 
    \includegraphics[width=2.7 in ]{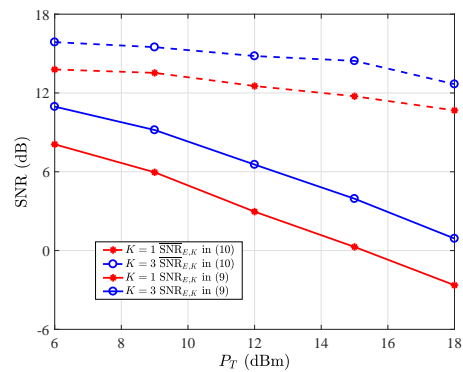}}
  \caption{ Wiretapping performance when there is no detection at the BS.}
  \label{NoDetection} 
  \vspace{-5 mm}
\end{figure}

In Fig. \ref{NoDetection}, we evaluate the SNR of the target Eve under the condition that the BS is unaware of the PSA.

In Fig. \ref{SchemeCom}, we compare the performances of the cooperative attacking scheme (CAS) proposed in this paper with a non-cooperative attacking scheme (NCAS).
In the NCAS, each Eve transmits with the maximal power by setting $\nu_k=\sqrt{P}$ for $k=1,2,\cdot\cdot\cdot,K$ without cooperating with the other Eves to optimize the PSA.
The results are obtained by averaging w.r.t. 1000 channel realizations.
For each channel realization, the SNR of the NCAS  is obtained by choosing the maximal wiretapping SNR across all Eves.
From Fig. \ref{SchemeCom}, we can observe that cooperation among Eves leads to significantly higher wiretapping SNRs.

In Fig. \ref{SumNoDetectionSNRvsPt}, the wiretapping SNR in \eqref{SNREk}, where the Eves do not have knowledge of $\pmb{h}_B$, and its upper bound in \eqref{SNREkUpperBound}, where $\pmb{h}_B$ is known by the Eves, are plotted.
As can be observed, the wiretapping SNR decreases with increasing $P_T$,
since a higher training power makes the channel estimation procedure more robust against the PSA.
However, we also observe that increasing $P_T$ can not significantly reduce the wiretapping performances when $\pmb{h}_B$ is known by the Eves. Therefore, from the perspective of secure transmission, the legitimate links should always keep their CSIs secret from the Eves.

\subsection{PSA Against Detection}
\begin{figure}
    \centering
  \subfigure[The general case]{
    \label{SumAgainstDetectionGeneral} 
    \includegraphics[width=2.7 in]{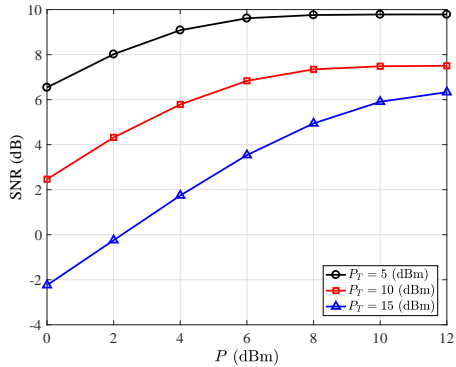}}
  \subfigure[The worst case]{
    \label{SumAgainstDetectionWorst} 
    \includegraphics[width=2.7 in]{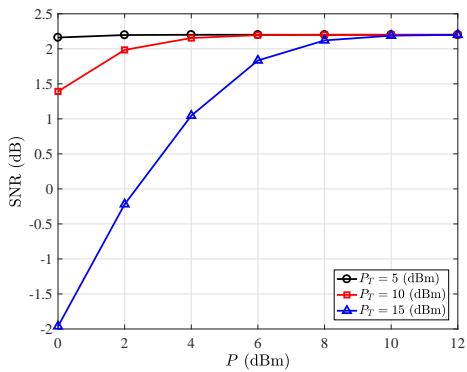}}
  \caption{The wiretapping SNR versus the attack power constraint $P$, where we set the upper bound on the detection probability to $\epsilon=0.2$ for the general case in (a) and to $\epsilon=0.4$ for the worst case in (b).}
  \label{SumAgainstDetection} 
  \vspace{-5 mm}
\end{figure}

\begin{figure}
    \centering
  \subfigure[The general case]{
    \label{SumAgainstDetectionGeneral2} 
    \includegraphics[width=2.7 in]{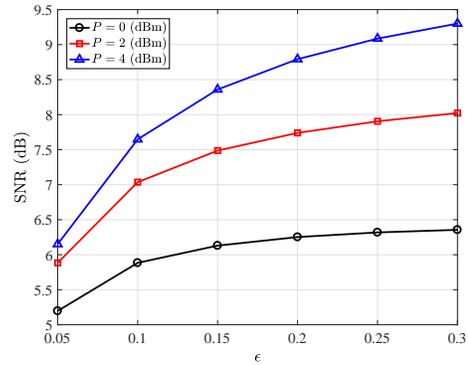}}
  \subfigure[The worst case]{
    \label{SumAgainstDetectionWorst2} 
    \includegraphics[width=2.7 in]{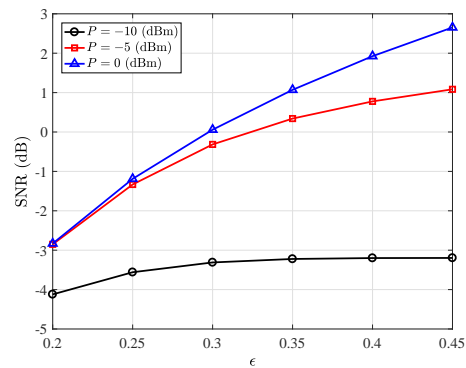}}
  \caption{The wiretapping SNR versus the upper bound on the detection probability $\epsilon$, where we set $P_T=5$ dBm.}
  \label{SumAgainstDetection2} 
  \vspace{-5 mm}
\end{figure}

Now, we assume the BS attempts to detect the PSA and the Eves design the PSA accordingly.
The performances of the attacking scheme under the general case, where the BS does not know $\pmb{h}_E$, and the worst case,  where the BS knows $\pmb{h}_E$, are simulated in Fig. \ref{SumAgainstDetection} and Fig. \ref{SumAgainstDetection2}, respectively.
We set $\eta_G = \eta_W =0.05$, $P_S=20~\mathrm{dBm}$ , $N=8$, and $K=3$.

In Fig. \ref{SumAgainstDetection}, we show the wiretapping SNRs versus the attacking power constraint for different training powers $P_T$.
As can be observed from Fig. \ref{SumAgainstDetection},  for all cases, for high $P$, the wiretapping SNRs become saturated.
This is because higher attacking powers make the PSA easier to detect by the BS, and when the attacking power is sufficiently high, the wiretapping performance is limited by the detection probability.

Fig. \ref{SumAgainstDetection2} shows the wiretapping SNR as a function of the upper bound on the detection probability for different attacking power constraints.
From Fig. \ref{SumAgainstDetection2}, we observe a trade-off between the wiretapping performance and the risk of the PSA of being detected. If a higher wiretapping SNR is desired, the risk of being detected increases. On the contrary, if the Eves want to avoid detection of the PSA, then the wiretapping performance will suffer.

\section{Conclusion}
In this paper, we investigated the case where the PSA is carried out by multiple Eves.
We assumed that the Eves collaborate during the training phase to maximize the wiretapping SNR of a target Eve and studied two different scenarios:
(1) the BS is unaware of the PSA,
and
(2) the BS attempts to detect the PSA.
For both scenarios, wiretapping SNR maximization problems were formulated and unified into one general non-convex optimization problem. An efficient MM-ADMM-based algorithm was developed to solve the general problem.
Our simulation results reveal that:
a) the MM-ADMM algorithm achieves near-optimal performance,
b) the cooperation among Eves can significantly improve their wiretapping capability,
c) the CSIs of the legitimate links should be kept secret from the Eves, otherwise, high wiretapping SNRs can be attained even for low attacking powers, and
d) if the Eves desire a higher wiretapping performance, they have to take a higher risk of being detected.


\appendix
\subsection{Brief introduction to the MM method}
\label{MMIntroduction}
For a general optimization problem $\mathop{\mathrm{max}}\limits_{\pmb{x}\in \mathcal{X}}~f(\pmb{x})$, where $\mathcal{X}$ is a convex set, the objective function may be in a complicated form, which makes the problem intractable. The MM method deals with this challenge by iteratively solving $\pmb{x}_n=\mathop{\mathrm{max}}_{\pmb{x}\in \mathcal{X}}~g\left(\pmb{x};\pmb{x}_{n-1}\right)$ starting from a feasible initial point $\pmb{x}_0$, where $g\left(\pmb{x};\pmb{x}_{n-1}\right)$ is an approximate function of $f\left(\pmb{x}\right)$ and $\pmb{x}_n$ is the solution obtained in the $n^{\mathrm{th}}$ iteration. If $g\left(\pmb{x};\pmb{x}_{n-1}\right)$ satisfies $g\left(\pmb{x};\pmb{x}_{n-1}\right)\leq f\left(\pmb{x}\right)$,
$g\left(\pmb{x}_{n-1};\pmb{x}_{n-1}\right) = f\left(\pmb{x}_{n-1}\right)$ and $\frac{\partial g\left(\pmb{x};\pmb{x}_{n-1}\right)}{\partial\pmb{x}}|_{\pmb{x}=\pmb{x}_{n-1}}=\frac{\partial f\left(\pmb{x}\right)}{\partial\pmb{x}}|_{\pmb{x}=\pmb{x}_{n-1}}$ for any positive integer $n$, then the sequence $\left\{f\left(\pmb{x}_n\right)\right\}_{n=1}^{+\infty}$ is monotonically increasing and finally converges to a stationary point of the original problem. In general, $g\left(\pmb{x};\pmb{x}_{n-1}\right)$ is chosen to have a simpler form than $f(\pmb{x})$, which makes the iterations efficient.
For more details about the MM method, please refer to \cite{MMmethod1,MMmethod2} and references therein.
\subsection{Brief introduction to ADMM algorithm}
\label{ADMMMMIntroduction}
The ADMM algorithm is used to deal with convex problems of the following form,
\begin{align}
\label{ReviewADMM}
\mathop{\mathrm{min}} \limits_{\pmb{x}\in \mathcal{X}; \pmb{z}\in \mathcal{Z}} \quad f\left(\pmb{x}\right) + g\left(\pmb{z}\right), \quad\quad \mathrm{s.t.} \quad \pmb{G}\pmb{x} = \pmb{z},
\end{align}
where $f\left(\pmb{x}\right)$ and $g\left(\pmb{z}\right)$ are convex functions and $\mathcal{X}$ and $\mathcal{Z}$ are non-empty convex sets.
The ADMM solves the above optimization problem by iterating the following updates,
\begin{align}
\pmb{x}^{(n)} &= \mathop{\mathrm{argmin}}\limits_{\pmb{x}\in \mathcal{X}}
\left\{\begin{aligned}
&f\left(\pmb{x}\right)
+\frac{\rho}{2}\left\|\pmb{G}\pmb{x}-\pmb{z}^{(n-1)}\right\|^2 \\
&+\Re\left\{\left\langle \pmb{y}^{(n-1)},\pmb{G}\pmb{x}-\pmb{z}^{(n-1)}\right\rangle \right\}
\end{aligned}\right\},\nonumber\\
\pmb{z}^{(n)} &= \mathop{\mathrm{argmin}}\limits_{\pmb{z}\in \mathcal{Z}}
\left\{\begin{aligned}
&g\left(\pmb{z}\right) + \frac{\rho}{2}\left\|\pmb{G}\pmb{x}^{(n)}-\pmb{z}\right\|^2\\
&+ \Re\left\{\left\langle \pmb{y}^{(n-1)},\pmb{G}\pmb{x}^{(n)}-\pmb{z} \right\rangle\right\}
\end{aligned}\right\},\nonumber\\
\pmb{y}^{(n)} &= \pmb{y}^{(n-1)} + \rho\left(\pmb{G}\pmb{x}^{(n)} - \pmb{z}^{(n)}\right),\nonumber
\end{align}
where $\rho$ is any positive real number and $\{\pmb{x}^{(n)},\pmb{z}^{(n)},\pmb{y}^{(n)}\}$ denotes the solution in the $n^{\mathrm{th}}$ iteration.
According to \cite[Proposition 4.2]{D.P.Bertisekas1989}, if the optimal solution set of the original problem \eqref{ReviewADMM} is not empty and the matrix $\pmb{G}^H\pmb{G}$ is invertible, then $\pmb{x}^{(+\infty)}$ is an optimal solution of \eqref{ReviewADMM}.
For more details about the ADMM alogrithm, please refer to \cite{S.Boyd2011,D.P.Bertisekas1989,K.Huang2016,E.ChenTCOM2017} and references therein.

\subsection{The proof of Proposition 1}
\label{ProofP}
{
Similar proofs for the optimality of SDR-based methods can be found in, e.g., \cite{Z.q.Luo,L.N.Tran,W.C.Liao} and references therein. Therefore, we only show a sketch of the  proof for the problem at hand.

The KKT conditions for \eqref{CounterDetectionSDP4Object} are as follows
\begin{subequations}
\label{FinalKKT}
\begin{align}
&\pmb{\Theta} - \mathrm{diag}\left\{\varphi_1,\varphi_2,\cdots,\varphi_K\right\} - \vartheta\pmb{T} + \omega\pmb{T} + \pmb{Z}= 0,\label{KKT1} \\
&\sum\nolimits_{i=1}^K\varphi_iP_i + \vartheta\varpi^2 + \omega\varrho = 0, \label{KKT2}\\
&\pmb{Z}\pmb{X} = \pmb{X}\pmb{Z} = \pmb{0},\quad\pmb{Z}\succeq\pmb{0},\quad\pmb{X}\succeq\pmb{0},\label{KKT3}\\
&~\vartheta\geq0,\quad \mu_i\geq0,\quad i=1,2,\cdot\cdot\cdot,K,
\end{align}
\end{subequations}
where $\left\{\varphi_i\right\}_{i=1}^{K}$, $\vartheta$, $\omega$, and $\pmb{Z}$ are the dual variables for constraints \eqref{CounterDetectionSDP3C1}, \eqref{CounterDetectionSDP3C2}, \eqref{CounterDetectionSDP3C3}, and the constraint $\pmb{X}\succeq \pmb{0}$, respectively.
From \eqref{KKT1}-\eqref{KKT3}, we obtain
$\pmb{Z} = \sum_{i=1}^K\varphi_i\pmb{D}_i + \vartheta\pmb{T} + \frac{\sum_{i=1}^K\varphi_iP_i + \vartheta\varpi^2}{\varrho}\pmb{T} - \pmb{\Theta}\succeq\pmb{0}$.
According to random matrix theory, $\pmb{T}=\pmb{A}^H\pmb{A}$ is a full rank matrix with probability $1$ when $K\leq N$.
Therefore,
$\mathrm{rank}\left(\pmb{Z}\right)\geq K-1$ with probability $1$ since $\mathrm{rank}\left(\pmb{\Theta}\right)=1$.
Due to the fact that $\pmb{Z}\pmb{X} = \pmb{X}\pmb{Z} = \pmb{0}$, we obtain that  $\mathrm{rank}\left(\pmb{X}\right)\leq 1$. Therefore, for non-zero $\pmb{X}$, we have $\mathrm{rank}\left(\pmb{X}\right)= 1$.
}

\end{document}